\documentclass[11pt]{article}
\usepackage[totalwidth=385pt,totalheight=584pt]{geometry}
\usepackage{amsmath,amssymb,epsf,graphics,graphicx}
\usepackage{bbm}
\usepackage{psfrag}
\newcommand\CPT{ {\cal CPT}}
\newcommand\CH{ {\cal H}}

\newcommand{\CM}{{\cal M}}

\newcommand{\One}{\mathbbm{1}}
\newcommand{\ZZ}{\mathbb{Z}}

\newcommand\blank[1]{}
\newcommand{\fract}[2]{{\textstyle\frac{#1}{#2}}}

\renewcommand{\hat}{\widehat}
\newcommand\eq{\begin{equation}}
\newcommand\en{\end{equation}}
\newcommand\bea{\begin{eqnarray}}
\newcommand\eea{\end{eqnarray}}
\newcommand\nn{\nonumber}
\newcommand\ba{\(\begin{array}}
\newcommand\ea{\end{array}\)}
\newcommand{\resection}[1]{\setcounter{equation}{0}\section{#1}}

\newcommand{\Z}{{\mathbb Z}}
\newcommand{\NN}{{\mathbb N}}

\newcommand\Rth{{\mathbb R}}
\newcommand\p{{\sf p}}
\newcommand\q{{\sf q}}

\newcommand\PT{{\cal P}{\cal T}}
\newcommand{\RR}{{\hbox{$\rm\textstyle I\kern-0.2em R$}}}
\begin{document}
\begin{titlepage}
\vskip 0.5cm
\begin{flushright}
DCPT-09/35
\end{flushright}
\vskip 1.2cm
\begin{center}
{\Large{\bf From $\PT$-symmetric quantum mechanics to conformal field theory }}
\end{center}
\vskip 0.8cm \centerline{Patrick Dorey$^{1}$,
Clare Dunning$^2$,
and Roberto Tateo$^3$} \vskip 0.9cm
\centerline{${}^1$\sl\small Dept.\ of Mathematical Sciences,
Durham University,} \centerline{\sl\small Durham DH1 3LE, UK}
\vskip 0.3cm \centerline{${}^{2}$\sl\small IMSAS, University of
Kent, Canterbury CT2 7NF, UK}
\vskip 0.3cm \centerline{${}^{3}$\sl\small Dip.\ di Fisica Teorica
and INFN, Universit\`a di Torino,} \centerline{\sl\small Via P.\
Giuria 1, 10125 Torino, Italy}

\vskip 0.2cm \centerline{E-mails:}
\centerline{p.e.dorey@durham.ac.uk, t.c.dunning@kent.ac.uk,
tateo@to.infn.it}

\vskip 1.25cm
\begin{abstract}
\noindent
One of the simplest examples of a $\PT$-symmetric quantum system is
the scaling Yang-Lee model, a quantum field theory with cubic
interaction  and purely imaginary coupling.
We give a historical review of some
facts  about  this model in $d \le 2$ dimensions,
from its original definition in connection with phase transitions
in the Ising model and its relevance to polymer physics, to
the r\^ole it has played  in studies of integrable quantum field
theory and of $\PT$-symmetric quantum mechanics.  We also
discuss  some more general results on  $\PT$-symmetric quantum
mechanics and the ODE/IM correspondence, and mention applications
to magnetic systems and  cold atom physics.
\end{abstract}
\end{titlepage}
\setcounter{footnote}{0}
%%%%%%%%%%%%%%%%%%%%%%%%%%%%%%%%%%%%%
\def\thefootnote{\fnsymbol{footnote}}
%%%%%%%%%%%%%%%%%%%%%%%%%%%%%%%%%%%%%%%%%%%%%%%%%%%%%%%%%%%%%%%%%%%%%
%%%  start of the paper  %%%%%%%%%%%%%%%%%%%%%%%%%%%%%%%%%%%%%%%%%%%%
%%%%%%%%%%%%%%%%%%%%%%%%%%%%%%%%%%%%%%%%%%%%%%%%%%%%%%%%%%%%%%%%%%%%%
%
\resection{Introduction}

The scaling Yang-Lee model in two dimensions is one of the   simplest
examples
of an interacting  integrable quantum field theory. It has often been  used
by the integrable models community as a testing ground
for  new mathematical and numerical tools.
Aspects of the theory of exact S-matrices~\cite{Cardy:1989fw, Smirnov:1989hh},
the thermodynamic Bethe Ansatz (TBA)~\cite{Zamolodchikov:1989cf},
the truncated conformal space approximation (TCSA)~\cite{Yurov:1989yu,
  Dorey:1997yg},
the form factor approach for  correlation
functions~\cite{Zamolodchikov:1990bk, Dorey:2000eh},
the exact off-critical  g-function~\cite{LeClair:1995uf,Dorey:2004xk},
the excited state and  the boundary TBA
methods~\cite{Bazhanov:1996aq,Dorey:1996re, LeClair:1995uf,Dorey:1997yg}
are all techniques which were
very successfully applied in their  early stages to the
study of non-perturbative phenomena in this simple model.

Deep in the ultraviolet regime the properties of the  scaling Yang-Lee
model are governed by the  conformal field theory
$\CM_{2,5}$~\cite{Cardy:1985yy, DiFrancesco:1997nk} which has a
negative central charge and a single relevant spin-zero field $\phi$,
with negative conformal dimensions. The  off-critical integrable
version of the model corresponds to the perturbation of the
conformally-invariant $\CM_{2,5}$ action by the operator $\phi$ with a
purely imaginary coupling constant~\cite{Cardy:1989fw}.

Despite its apparent lack of unitarity, the Yang-Lee model in
$d$ space-time dimensions is at least relevant in condensed matter
physics: it is related to  the theory of non-intersecting branched
polymers in $d+2$ dimensions~\cite{Parisi:1980ia}.

Considerations of the Yang-Lee model led
Bessis and Zinn-Justin to conjecture the reality of the
spectrum of the Schr\"odinger equation with cubic potential and
purely-imaginary coupling, which in turn prompted Bender and his
collaborators to a more general study of Schr\"odinger problems with complex
potentials~\cite{Bender:1998ke,Bender:1998gh}. The
paper~\cite{Bender:1998ke} marks the beginning of $\PT$-symmetric quantum
mechanics  and quantum field theory as an area of intensive research.
This short review and the corresponding
introductory lecture given in Mumbai are especially addressed to
students approaching the field of quantum mechanics  in the complex
domain. However,  the brief  discussions on  conformal field theory,
condensed matter physics, integrable models and their correspondence
with the theory of ordinary differential equations may also stimulate
more experienced researchers to look into  some of the many  open
problems related to non-unitary integrable quantum field theories.
\resection{ The Yang-Lee edge singularity }
\label{sec1}
Consider the  ferromagnetic  Ising model with non-zero external
magnetic field $\text{H}$
\eq
\CH[{\bf s},\text{H}]=  -\sum_{<ij>} s_i s_j + \text{H} \sum_i s_i~,~~~s_i=\pm 1
\en
where the sum is over nearest-neighbour sites in a $2d$ square lattice. The partition function is
\eq
Z(T,\text{H}) = \sum_{s_i=\pm 1} e^{-\CH[{\bf s},\text{H}]/T}~.
\en
In the thermodynamic limit, at $\text{H}=0$ with $T$ above a certain critical
temperature $T_c$,  the system is in
a  disordered phase  with zero magnetisation
$\text{M}(T,\text{H})|_{\text{H}=0} =\langle s_i \rangle =0$. At  $T<T_c$ the system  is
instead  in an ordered phase
where the $\Z_2$ symmetry is spontaneously broken and  $\text{M}(T,\text{H})|_{\text{H}=0}
\ne 0$. Finally at $\text{H}=0$ and  $T=T_c$  the model undergoes a   second
order phase transition where typical spin configurations exhibit
fluctuations at all length scales
and the  continuum  limit version of the model is invariant under
conformal transformations.

A result that dates back to Yang and Lee in
1952~\cite{Yang:1952be,Lee:1952ig}
states that the partition function $Z(T,\text{H})$  of the finite lattice version of
the Ising model, at fixed  $T$,
is an entire function of $\text{H}$ with all the zeros located on the purely
imaginary $\text{H}$-axis.  In the thermodynamic limit the zeros become
dense,
covering
the entire imaginary  axis  from $-\infty$ to
$\infty$ apart from a possible
single finite gap centered at $\text{H}=0$.
When $T<T_c $, there is no gap in the zero distribution and the
magnetization  $\text{M}(T,\text{H})$
has a finite discontinuity  as  $\text{H}$ crosses the origin along the real
axis. Correspondingly the system undergoes a first-order phase
transition.

\begin{figure}[ht]
\begin{center}
{~}\qquad\qquad\qquad\qquad
\epsfxsize=.585\linewidth\epsfbox{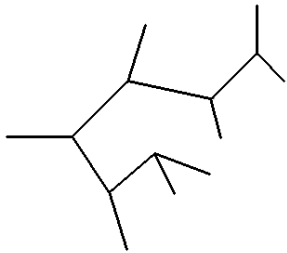}
\end{center}
\vspace{-25pt}
\caption{\small A branched-polymer tree diagram.} \label{figfeynman}
\end{figure}

For $T>T_c$ there is a gap  for
$|\text{H}|<h_c(T)$.  The edges $\text{H}=\pm ih_c(T)$  of this gap are  branch points for the magnetization  $\text{M}(T,\text{H})$
and according to  Fisher~\cite{Fisher:1978pf}   they  can  be considered as  conventional critical points.
In higher dimensions they  correspond to the infrared behavior of a scalar field theory  with  action
\eq
A= \int d^dx~ \left(\frac{1}{2} (\nabla \phi)^2 + i(h-h_c) \phi
+i\frac{g}{3} \phi^3~\right)~.
\label{action}
\en
The action (\ref{action}) is related to the   Parisi-Sourlas theory
of non-intersecting branched polymers in $D=d+2$
dimensions~\cite{Parisi:1980ia}.
\resection{The Yang-Lee model and branched polymers}
\label{brpoly}
Following~\cite{Parisi:1980ia, Cardy:2001ci, Cardy:2001ci1}, non-intersecting
branched polymers in $D$ dimensions correspond
to the classical field equation
\eq
\nabla^2_D \phi(x)+V'(\phi)+i\xi(x)=0~,
\label{classical}
\en
where the field potential  $V(\phi)$ is
\eq
V(\phi)= -(h-h_c) \phi(x) + \frac{g}{3} \phi^3(x)~,
\en
and  $\xi(x)$  is a stochastic (white noise) variable with
\eq
\langle \xi(x) \xi(y)  \rangle = \delta^D(x-y).
\en
The polymers can be  visualized  using  tree Feynman diagrams  (see
figure~\ref{figfeynman})
with  corresponding Feynman rules:
\begin{itemize}
\item{A segment in the  graph  corresponds to the  Green's function of
$-\nabla^2_D$~.}
\item{ Each vertex carries a factor $g$~.}
\item{ Each free end carries a factor $-(h-h_c)+i \xi$. }
\end{itemize}
The $n$-point correlation functions are defined as
\eq
\langle  \phi(x_1) \dots \phi(x_n) \rangle= \int {\cal D} \xi ~
\phi(x_1)\dots \phi(x_n) e^{ - \frac{1}{2} \int dz^D \xi^2(z)}~,
\en
where the $\phi(x_i)$'s solve the classical field equation (\ref{classical}).

Resolving the constraints using  the Faddeev-Popov  method
introduces  a Grassmann ghost field  and reveals that
the theory has an unexpected  invariance under supersymmetric
transformations.
A consequence of this hidden supersymmetry  is a surprising
dimensional reduction property: correlation functions whose arguments
are restricted to a $d=D-2$ subspace are the same as those for the
Yang-Lee theory in $d$ dimensions.

The simplest case corresponds to the branched polymer problem in
$D=2$ dimensions.
In  $d=D-2=0$ the  gradient term in the Yang-Lee action~(\ref{action}) is absent and
the vacuum-vacuum
functional integral  becomes,  after a complex rotation,
\eq
\frac{2 \pi}{g^{1/3}} {\rm Ai}\left(\frac{(h-h_c)}{g^{1/3}} \right) =\int_{-\infty}^{\infty}  e^{i(h-h_c) \phi + i\frac{g}{3} \phi^3 } d\phi~.
\label{airy}
\en
This is one of the very few exactly-known
scaling functions in $D=2$. This model describes  the crossover from
self-avoiding
loops (vesicles) to branched polymers~\cite{Richard:2001sb,Cardy:2001ci,Cardy:2001ci1}
where $p \sim g^{-1}$ is the  pressure difference between the inside and the outside
and  $(x_c-x) \sim (h-h_c) g^{-1}$ is related to the monomer  fugacity in the  discretized version of the model.
\begin{figure}[ht]
\begin{center}
\epsfxsize=.585\linewidth\epsfbox{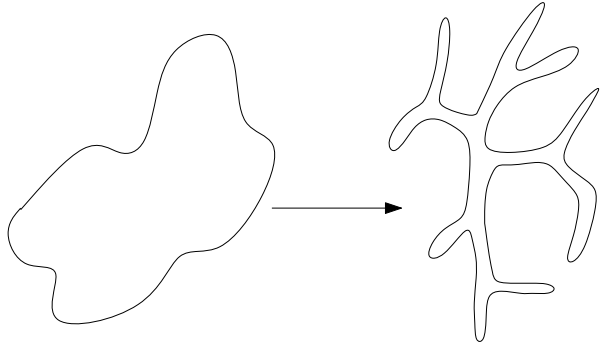}
\end{center}
\caption{\small       The self-avoiding loops to branched polymers
crossover transition.} \label{fig3}
\end{figure}
This result can be interpreted from another  perspective:
since the original polymer problem is defined in $D=2$ dimensions  and
the  Airy function is the solution
of a $1d$ Schr\"odinger equation with linear potential, we have a
beautiful  physically-motivated
confirmation  of an ODE/IM correspondence~\cite{Dorey:1998pt}
linking the spectral theory of simple  ordinary differential equations
and the theory of integrable models.
\resection{Conformal field theory in two dimensions:
the Ising and the Yang-Lee models}
From section~\ref{sec1} we know that  the Ising model at  $\text{H}=0$ and
$T=T_c$ is critical and invariant under conformal transformations.
The edge points  $\text{H}=\pm ih_c(T)$  can also
be considered  as  conventional critical points~\cite{Fisher:1978pf} where the model
exhibits conformal invariance~\cite{Cardy:1985yy}. From a renormalisation group
perspective  the point $(\text{H}=0, \ T=T_c)$ and the pair of equivalent points
$(\text{H}=\pm ih_c(T),\ T>T_c)$ correspond to  two  distinct universality
classes.

Two dimensional  conformal transformations coincide with  analytic
transformations
\eq
z \rightarrow f(z)~~,~~\bar{z} \rightarrow \bar{f}(\bar{z})
\label{conf}
\en
with $z=x+iy $, $\bar{z}=x-iy $. The corresponding infinitesimal generators are
\eq
l_n= - z^{n+1} \partial_z~~,~~\bar{l}_n= - \bar{z}^{n+1} \partial_{\bar{z}}~~,~~
\en
and satisfy the Witt algebra
\eq
[l_m,l_n]=(m-n) l_{m+n}~,~[\bar{l}_m,\bar{l}_n]=(m-n)
\bar{l}_{m+n}~,~[l_n,\bar{l}_m]=0~.
\en
At the quantum level the geometry is important  and commutation relations
involving infinitesimal generators
that change it  become anomalous. The Witt algebra is
replaced by  the Virasoro algebra
$l_m \rightarrow L_m $ , $\bar{l}_m \rightarrow \bar{L}_m $,
where $[L_n,\bar{L}_m]=0 $ and
\bea
{}[L_m,L_n]&=&(m-n) L_{m+n}+  {c \over 12} (n^3-n)
\delta_{n+m,0}~,\nn\\[3pt]
{}[\bar{L}_m,\bar{L}_n]&=&(m-n) \bar{L}_{m+n}+
{c \over 12} (n^3-n) \delta_{n+m,0}~,
\label{virasoro}
\eea
with  $c$ the Casimir coefficient  describing the coupling of the
quantum fluctuations to the  finite geometry of the system. The
quantity $c$ is also called the central charge, or conformal anomaly.
%
%%%%%%%%%%%%%%%%%%%%%%%%%%%%%%%%%%%%%%%%%%%%%%%%%%%%%%%%%
\subsection{ The Kac table}
In two dimensions a universality class is completely identified by
its central charge and a complete set of
operators, or fields, with their  operator product expansions.
Among all the operators the  primary operators play a central r\^ole:
they are the ones that  transform  in the
following simple way under a conformal   transformation (\ref{conf})
\eq
\phi(z,\bar{z}) \rightarrow
(\partial_z f(z))^h (\partial_{\bar{z}} \bar{f}(\bar{z}))^{\bar{h}}
\phi(f(z),\bar{f}(\bar{z}))~,
\en
where  $h$ and $\bar{h}$ are the holomorphic
and the anti-holomorphic conformal dimensions
respectively\,\footnote{ In the
  following we are only interested in spin-zero operators with
  $h=\bar{h}$.}.
The remaining operators -- called descendants --
can be obtained from the primary operators by
acting on them with combinations of
 $L_{-n}$ and $\bar{L}_{-n}$.
For generic values of $c$ there are
infinitely-many primary operators, but in their pioneering paper
\cite{Belavin:1984vu}
on conformal field theory, Belavin, Polyakov and Zamolodchikov showed
that for each pair of
coprime integers $\p < \q $, a `minimal' conformal field theory
$\CM_{\p,\q}$ can be defined, with
\eq
c=1- 6 {(\p-\q)^2  \over \p \q}<1
\en
and a {\em finite}\/
number of primary operators, with  conformal dimensions
\eq
h_{r,s}= { (\p s -\q r)^2 -(\p-\q)^2 \over 4 \p \q}~,~~~(1 \le r < \p,
\ 1 \le s < \q r/\p~)~.
\en
The set of conformal dimensions  forms the so-called Kac table of the model.
The critical Ising model corresponds to the minimal model   $\CM_{3,4}$  with
central charge $c=1/2$ and Kac table
\eq
\One \leftrightarrow h_{1,1}=\bar{h}_{1,1}=0~~,~~\epsilon \leftrightarrow h_{1,3}=\bar{h}_{1,3}=1/2~~,~\sigma \leftrightarrow
h_{1,2}=\bar{h}_{1,2}=1/16~,
\en
where $\One$ is the identity operator, $\epsilon$ and $\sigma$ are respectively the energy and the
magnetic field operators, while
the Yang-Lee edge singularity corresponds to  $\CM_{2,5}$\,, with central
charge $c=-22/5$ and Kac table
\eq
\One \leftrightarrow h_{1,1}=\bar{h}_{1,1}=0~~,~~\phi \leftrightarrow
h_{1,2}=\bar{h}_{1,2}=-1/5~.
\en
Notice that the second of these conformal dimensions is negative,
a reflection of the non-unitary nature of the Yang-Lee model.

\resection{The Scaling Yang-Lee model in two dimensions}
A $2d$ conformal field theory is a particular example of an integrable quantum
field theory,
the integrability of the model being  a direct consequence of the
existence of the infinite number of commuting conserved charges built
from $\{ L_n,
\bar{L}_n \}$. It was an idea of Sasha
Zamolodchikov~\cite{Zamolodchikov:1989zs} that a conformal field
theory can be perturbed or driven away from  the
critical point by one of the relevant spin-zero fields  of the
theory, in such a way that integrability would be preserved.
For particular primary fields he was able to deduce that
an infinite set of commuting conserved charges survives after  the
perturbation and  that  in these cases the model remains integrable.
It is now known that both
$\phi_{12}$ and $\phi_{13}$ perturbations of the minimal models lead
to integrable off-critical quantum field theories.
The perturbation of  $\CM_{2,5}$  by the  $\phi$ operator
\eq
A= A_{CFT}+ i (h-h_c) \int d^2 x \,\phi(x)
\label{apcft}
\en
is therefore integrable, and can be studied using powerful
techniques from the theory of integrable models. Results that
have been obtained include:
\begin{itemize}
\item{ The exact 2-body S-matrix  was proposed by Cardy and
Mussardo~\cite{Cardy:1989fw}
 and
independently by Smirnov~\cite{Smirnov:1989hh} in 1989.
The resulting theory for $(h-h_c)>0$ contains a single particle
species with a $\phi^3$ interaction. No other bound states exist.}
\item{ Based on the S-matrix
of~\cite{Cardy:1989fw,Smirnov:1989hh}, Aliosha Zamolodchikov
 derived
a set of  thermodynamic Bethe Ansatz (TBA) equations describing  the
exact ground-state energy of the model on an
cylinder with finite circumference~$R$~\cite{Zamolodchikov:1989cf}.}
\item{In 1996  the thermodynamic Bethe Ansatz  equations were
generalised to the full set of
excited  states~\cite{Bazhanov:1996aq, Dorey:1996re}. }
\item{ In 1990  Aliosha Zamolodchikov found an exact form-factor expansion  for particular
two-point  correlation functions~\cite{Zamolodchikov:1990bk}. }
\end{itemize}

\begin{figure}[ht]
\begin{center}
\epsfxsize=.43\linewidth\epsfbox{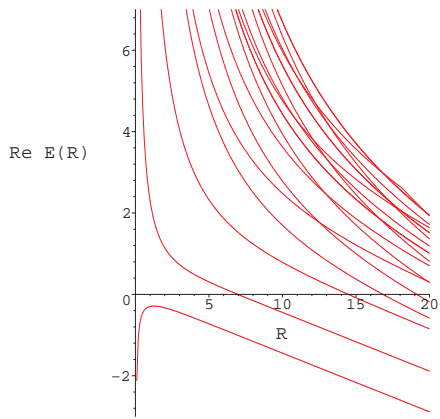}
\end{center}
\caption{\small Yang-Lee energy levels  for $(h-h_c)>0$. } \label{fig4}
\end{figure}

\begin{figure}[ht]
\begin{center}
{~}\!\!\!\!\!\!
\!\!\!\!\!\!
\epsfxsize=.44\linewidth\epsfbox{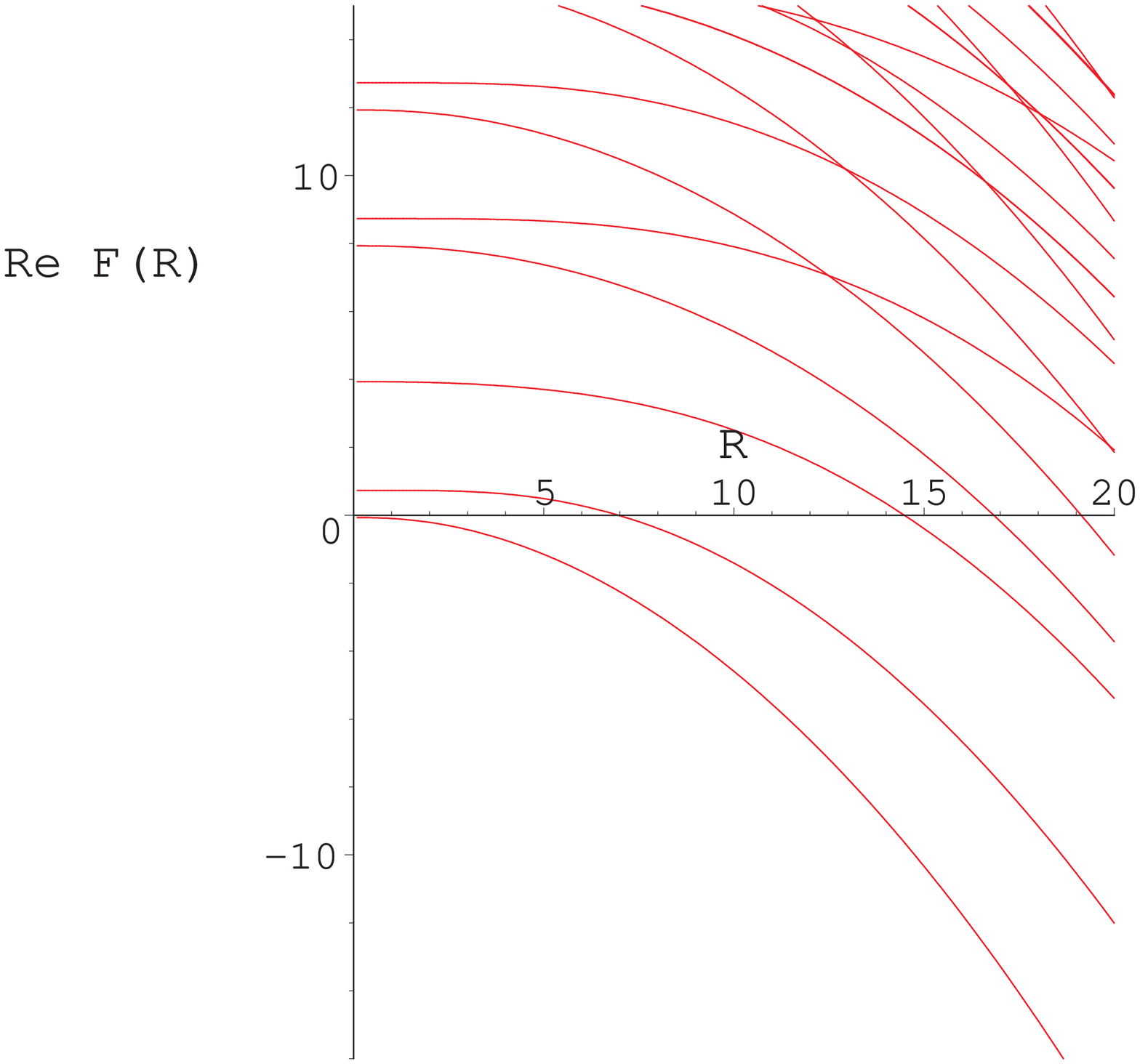}\qquad\qquad
\epsfxsize=.4\linewidth\epsfbox{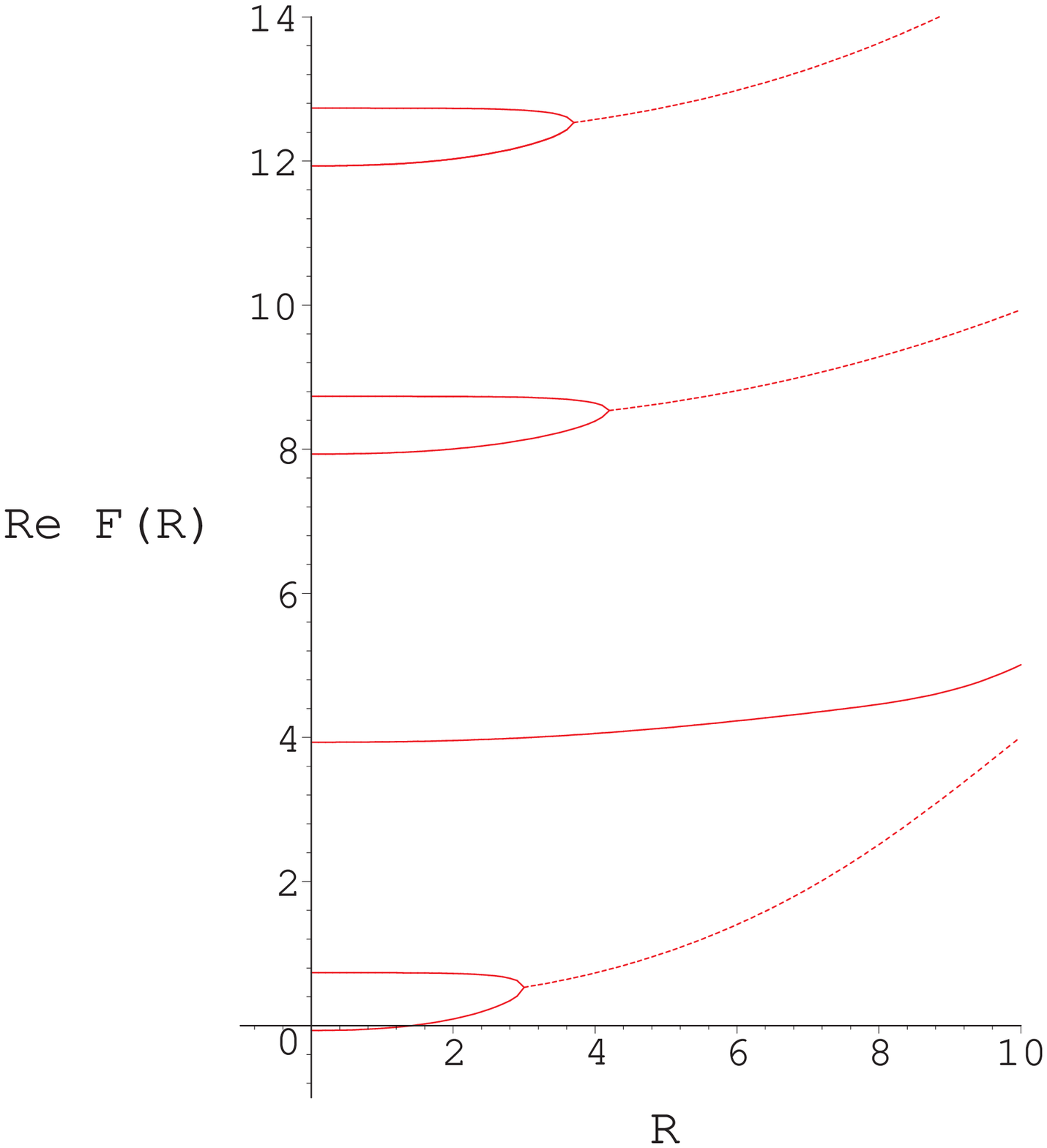}
\end{center}
\caption{\small  Yang-Lee scaling functions $F(R)=R E(R)/\pi$  for $(h-h_c)>0$
(left-hand plot) and $(h-h_c)>0$ (right-hand plot). } \label{fig5}
\end{figure}

Finally, the low-lying energy levels of the theory can be studied
numerically
using another method, proposed by  Yurov and Aliosha
Zamolodchikov~\cite{Yurov:1989yu},
called the truncated conformal space approximation.
This relies on the diagonalisation
of the perturbed Hamiltonian defined on a circle  of circumference $R$
on a suitable truncation of its Hilbert space, using the basis
provided by the (exactly-known) space of states of the unperturbed
conformal field theory.
Figures \ref{fig4} and \ref{fig5}  reproduce some of the results found by Yurov and
Zamolodchikov in 1990. Figure~\ref{fig4} shows the spectrum of the
Yang-Lee model  at fixed  $(h-h_c)>0$, as the circumference $R$ of the
cylinder  is increased.  One can see that the spectrum is completely
real. The
bottom line represents the ground-state energy $E_0$ and above
this line lies the single one-particle state $E_1$, the
energy gap  tending asymptotically to the mass $m$ of a single  neutral
particle at rest. Above this second line, asymptotically again at
$E_2-E_1=m$
there is the lowest  two-particle excited state, and above it the continuum.

\vskip -7pt

The situation for $(h-h_c)<0$ is very different.
For  small  values of $R$ all the energy levels are still real,  but
the ground and the first excited
states meet at a critical value $R_c$\,, beyond which they
form a complex-conjugate pair.
At the special point where this happens, there is a  square-root
branch point of the ground-state energy.

\vskip -7pt

Figure \ref{fig5} shows this and other
mergings, and illustrates the contrast between the
situations for the two signs of $(h-h_c)$. The plots show
the so-called scaling functions $F(R)\equiv R E(R)/\pi$.
Conformal field theory gives the (finite) limits of these functions
as $R\to 0$. For $R\neq 0$ the scaling functions  move away
from these limits
(in fact, as analytic functions of $(h-h_c)R^{12/5}$),
and for $(h-h_c)<0$ certain pairs
meet at square-root branch points, in a way that will see a striking
echo in a quantum-mechanical context in the next section.

As a more technical remark,
note that the explicit factor of $i$ appearing in the
action (\ref{apcft}) is a result of the particular
normalisation of the perturbing field $\phi(x)$ used
in~\cite{Yurov:1989yu}, which ensured that the corresponding
primary state in the conformal field theory had positive
norm-squared.
However there is no compelling reason to make such a choice:
if $L_n^{\dagger}=L_{-n}$, then it follows from the
Virasoro algebra (\ref{virasoro})
that, if $|\psi\rangle$ is the state corresponding to
a primary field $\psi(x)$, then the squared norm of, for example,
$L_{-n}|\psi\rangle$ is $2n(h+\frac{c}{24}(n^2{-}1))$ times that of
$|\phi\rangle$. In a non-unitary conformal field theory such as the
Yang-Lee model, where both $c$ and some of the conformal dimensions
$h$ are negative, there is
therefore no natural way for all (primary {\em and}\/ descendant)
states to have positive squared norms, and from many points of view it
is preferable to modify the normalisations chosen in
\cite{Yurov:1989yu}, so that all coefficients
in operator product expansions are real (see, for example,
\cite{Kausch:1996vq,Dorey:1997yg,Runkel:1998pm}).
This removes the factor of $i$
from (\ref{apcft}), but of course the theory remains non-Hermitian,
since the norm in the underlying Hilbert space is not
positive-definite. A similarly non-positive-definite norm
will appear in a
quantum-mechanical context in the next section.

Finally for this section, we should mention that $\PT$ symmetry can
also be found in integrable lattice models, where the continuous
dimensions of space and time are replaced by a discrete
lattice. An initial set of examples was discussed in
\cite{Korff:2007qg}, but it is clear that many further cases remain to
be studied.

%
%
%
%
%%%%%%%%%%%%%%%%%%%%%%%%%%%%%%%%%%%%%%%%%%%%%%%%%%%%%%%%%
%
\resection{ $\PT$-symmetric Quantum Mechanics }
\subsection{Spectral reality}
In the early 1990s, motivated by the Yang-Lee edge singularity,
Bessis and Zinn-Justin
 considered
the spectrum of the non-Hermitian Hamiltonian
\eq
H_{BZJ}=p^2 + ix^3
\label{cubic}
\en
defined on the full real line.
The
corresponding Schr\"odinger equation is
\eq
-\frac{d^2}{dx^2}\psi(x)+ix^3\psi(x)=E_n\psi(x) \quad , \quad \psi(x)
\in L^2 (\RR).
\en
Perturbative and numerical work led them to
conjecture \cite{BZJ}
\begin{itemize}
\item{The spectrum of $ H_{BZJ}$ is real, and positive.\footnote{While the
Bessis -- Zinn-Justin conjecture initiated the line
of work we want to describe here, similar observations had in fact
been made before. For example, in 1980 the
oscillators $p^2 + x^2 + igx^3$ were
studied~\cite{CGM}, and the reality of individual eigenvalues
proved for $g$ real and $|g|$ sufficiently small. However it is
important to note  that this
does {\em not}\/ suffice to prove that the whole spectrum is real for
any non-zero value of $g$. The reason is that the `sufficiently small'
value of $|g|$ may depend on the particular eigenvalue under
consideration. Without a global bound away from zero for these values,
reality can still be lost sufficiently high in
the spectrum for all non-zero values of $g$, even if each
individual eigenvalue ultimately becomes real as $|g|\to 0$.
(This is exactly the behaviour seen in Bender and Boettcher's
generalisation of the Bessis -- Zinn-Justin problem as $M\to 1^-$,
illustrated in figure \ref{f0} below, and it
is not ruled out by the arguments of \cite{CGM}).
A full proof of spectral reality for $p^2 + x^2 + igx^3$ -- in fact
valid for any real value of the parameter $g$ -- was finally provided
by Shin in 2002~\cite{Shin1}, following the approach
of~\cite{DDTb}\,.}}
\end{itemize}
In 1997,  Bender and Boettcher \cite{Bender:1998ke} proposed an
apparently-simple
generalization:
\eq
H_M=p^2-(ix)^{2M} ~,\qquad M \in \RR\,,~ M>0~.
\label{bb}
\en
To render the `potential' $ -(ix)^{2M} $ single-valued for all
values of $M$, a branch cut should be placed
along the positive imaginary axis of the complex $x$ plane.
For $M<2$, the analytic continuation of the spectrum at $M=3/2$
(Bessis and Zinn-Justin's original problem) can be recovered by
imposing the
boundary condition $\psi(x) \in L^2 (\RR)$, or equivalently
demanding that $\psi(x)$ should tend exponentially to zero as
$|x|\to\infty$ along the positive and negative real axes.
However, when  $ M $  reaches $ 2 $,  the  potential becomes  $-x^4$
and the eigenvalue problem  as just
stated changes its nature dramatically.
To see why, consider the
WKB approximation to the wavefunction as $x$ tends to infinity along
a general ray in the complex
plane:
\eq
\psi_{\pm}\sim \rho^{-M/2}\exp\left[\pm \fract{1}{M{+}1}e^{i\theta(1{+}M)}
\rho^{1{+}M}\right]~,~~~ (x=\rho e^{i \theta}/i)~.
\en
For most values of $ \theta $, one of these solutions
will be exponentially growing as $\rho\to\infty$
(a dominant solution)
and the other exponentially decaying (or subdominant).
But when $\theta$ is such that
\eq
\Re e[e^{i\theta(1{+}M)}]=0~,
\en
neither solution is dominant: both decay algebraically.
The angles
\eq
\theta=
\pm\frac{\pi}{2M{+}2}~,~
\pm\frac{3\pi}{2M{+}2}~,~
\pm\frac{5\pi}{2M{+}2}~,~\dots
\en
define anti-Stokes lines, which divide the complex plane
into Stokes sectors.  Across an anti-Stokes line, the dominant and
subdominant solutions swap r\^oles.
\begin{figure}[ht]
\begin{center}
\epsfxsize=.585\linewidth\epsfbox{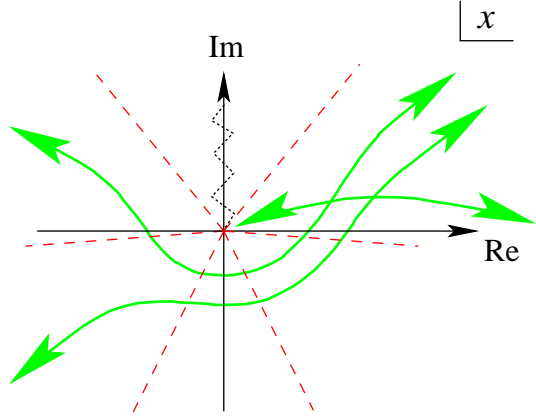}
\end{center}
\caption{\small  Sample quantisation contours. The dashed lines separate Stokes
  sectors.} \label{stokes}
\end{figure}

At  $M=2$, the positive and negative real axes
coincide with anti-Stokes lines,
and the original eigenproblem ceases to have a
discrete spectrum.  The issue can be resolved if the perspective is
widened by allowing the wavefunction to be defined along general
contours in the complex plane, rather than just the real axis.
Two types of eigenvalue problem are natural.
Lateral  problems correspond to boundary conditions which require
wavefunctions to decay at large $|x|$ along
a contour that begins and ends in a {\it pair}\/ of Stokes sectors.
By contrast, radial problems correspond to a
quantisation contour which starts at $x=0$, and runs out to infinity
in a given Stokes sector.
Figure \ref{stokes} depicts some sample quantisation
contours.  Some eigenproblems defined in this way will be related via
simple changes of variables, but those that are not have completely
different spectra.

\begin{figure}[ht]
\begin{center}
\epsfxsize=.585\linewidth\epsfbox{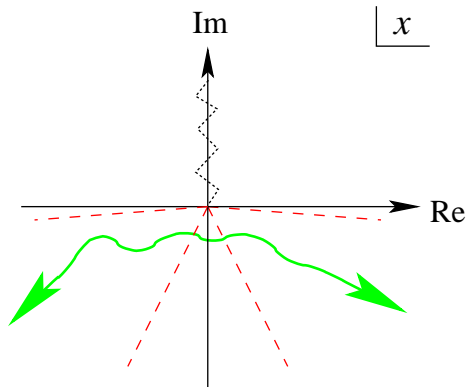}
\end{center}
\caption{\small The Bender-Boettcher
contour, for $M$ just larger than $2$.} \label{sectors}
\end{figure}

For $M<2$, the Bender-Boettcher  problem (\ref{bb}) is an
instance of the lateral
problem with
the (next-nearest-neighbour) Stokes sectors that
include the real axis. For $M>2$, these sectors rotate down in the
complex plane as shown in figure \ref{sectors}, and no longer include
the real axis.
Demanding  that the quantisation contour begins and ends inside this pair
of Stokes sectors for all positive values of $M$ gives the
analytic continuation of the
Bender-Boettcher  problems as $M$ increases past $2$.

Numerical and analytical evidence led Bender and Boettcher
to conjecture that, with these boundary conditions understood,  the
spectrum of (\ref{bb}) is real and positive for all $M>1$, despite the
non-hermiticity of the problem.   The real eigenvalues  as a function
of $M$ are shown in figure \ref{f0}. As soon as $M$ decreases
below $1$, infinitely-many eigenvalues pair off and
become complex, while the spectrum at $M=1/2$ is empty.

\begin{figure}[ht]
\begin{center}
\smallskip

\epsfxsize=.5\linewidth\epsfbox{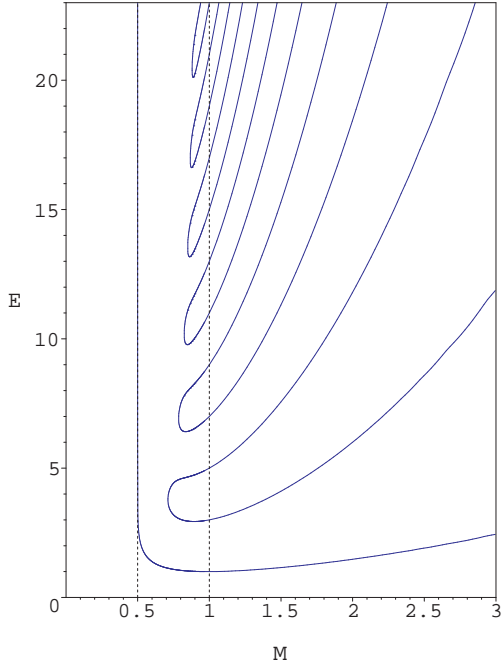}
%\qquad \qquad
\qquad \quad
\end{center}
\vskip -7pt
\caption{\small  Real eigenvalues of $p^2 -(ix)^{2M}$
  as a function of
  $M$.} \label{f0}
\end{figure}

For $M=2$,
the upside-down quartic potential along the
complex contour  described above,
the reality of the spectrum
 can be understood via
a spectrally-equivalent Hermitian
Hamiltonian \cite{Andrianov:1982,BG1993,jm2}.  Jones and Mateo  found a
particularly
transparent way to derive this result \cite{jm2} using
  the simple variable change
$x=-2i\sqrt{1+iw}$, which maps the complex contour  onto the full real
line.  This variable change followed by a Fourier transform turns the
$\PT$-symmetric Hamiltonian $H_2$ defined on the contour in
figure~\ref{sectors} into
\eq
{\widetilde H}_2 =p^2 +4w^4 - 2w
\en
defined on the full real line,
which is explicitly Hermitian.  Apart from $M=2$ and
$M=3$~\cite{DDTb}, mappings to Hermitian Hamiltonians
have not been found for the Bender-Boettcher problems  $H_M$ for
other values of $M>1$.

\begin{figure}[t]
\begin{center}
~\!\!\!\!\!\!
\!\!\!\!\!\!
\!\!\!\!\!\!
\!\!\!\!\!\!
\epsfxsize=.3\linewidth\epsfbox{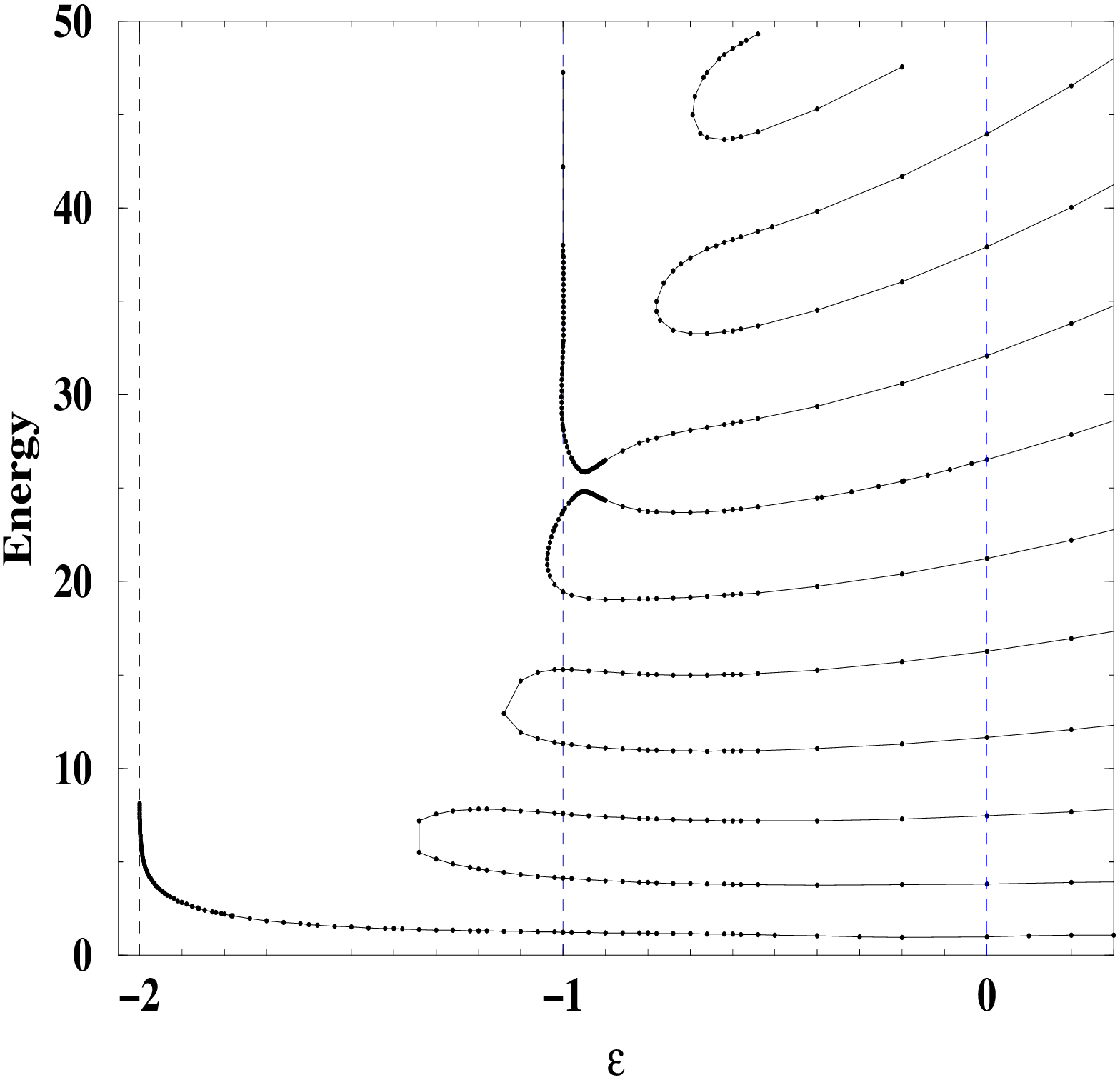}
{}~~~~~~\quad \qquad  \qquad \qquad
\epsfxsize=.3\linewidth\epsfbox{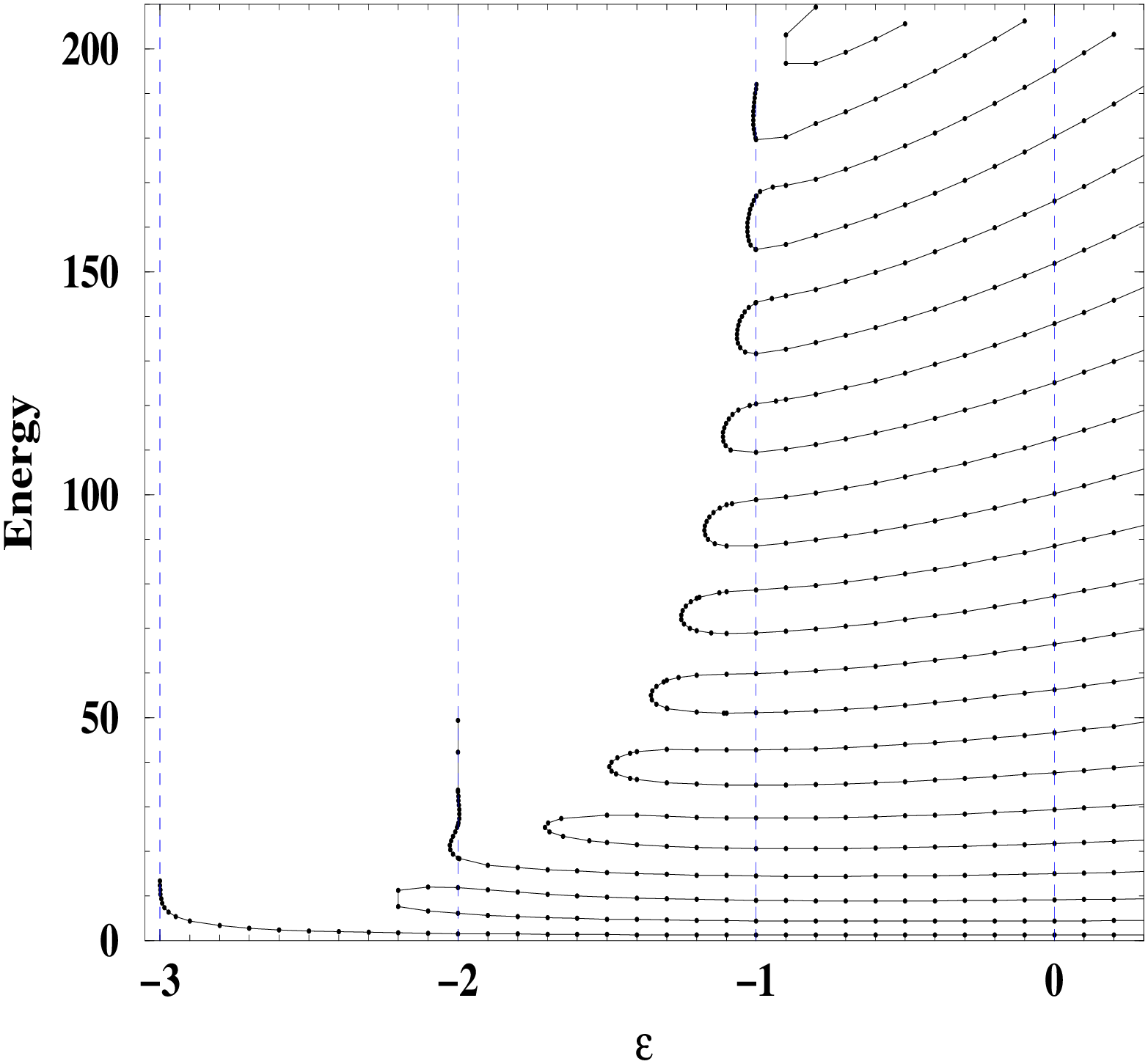}
\\[-9pt]
\parbox{0.33\linewidth}{~~~~~~\small\protect\ref{bb12}a: $K=2$}~~~~%&
{}~~~~\qquad \qquad \qquad
\parbox{0.33\linewidth}{~~~~~~\small\protect\ref{bb12}b:
$K=3$}
\end{center}
\vskip -5pt
\caption{\small\protect{ \label{bb12}}Real eigenvalues of
  $H^{(K)}_{\varepsilon}=p^2
  +x^{2K}(ix)^{\varepsilon}$ as a  function   of $\varepsilon$ (plots
taken from~\cite{Bender:1998gh}).
}
\end{figure}

%%%%%%%%%%%%%%%%%%%%%%%%%%%%%%%%%%%%%%%%%%%%%%%%%%%%%%%%%
%
%
Bender and Boettcher interpreted the   phase transition
from entirely real eigenvalues to infinitely-many complex
eigenvalues   as a spontaneous
breaking of $ {\cal PT} $ symmetry.  The  parity reflection ${\cal
  P}$ and time-reversal ${\cal T}$ operators act as follows:
\bea
{\cal P} &:&   x \rightarrow -x~,~p \rightarrow -p \nn \\
 {\cal T} &:&  x \rightarrow ~x~,~p \rightarrow -p~,~
i \rightarrow -i.\nn
 \eea
The combined operator ${\cal PT}$ commutes with the Hamiltonian
$H_M$ for all values of $M$.
The eigenvalues  of ${\cal PT}$ are pure
phases and  with  a multiplicative rescaling of the eigenfunctions they can be set equal to $1$.
If the eigenfunction $ \psi_n $ of $ H_M $
is also an eigenfunction of $ { \cal PT} $, then   we say that the
${\cal PT}$-symmetry  is unbroken.
In this case, using commutativity  we have
\eq
{\cal PT}\; H_M
\psi_n=  {\cal PT} \;E_n \psi_n = E^*_n \;{\cal PT} \psi_n=E^*_n \;  \psi_n
\en
\eq
H_M \;{\cal PT}\psi_n= H_M \; \psi_n=  E_n\psi_n
\en
and so $ E_n $  is real. Unfortunately, this does not seem to
offer a quick route to a proof of spectral reality, since it is
a non-trivial task to prove that $\PT$ symmetry is unbroken for all
wavefunctions. The only proof currently known
of the reality of the spectrum of $H_M$ for general $M>1$
\cite{DDTb}
was obtained by
a rather different route, using ideas from the ODE/IM correspondence
\cite{Dorey:1998pt}.

Spectra for problems involving
next-next-nearest-neighbour Stokes sectors
and next-next-next-nearest-neighbour Stokes sectors
are shown in figure
\ref{bb12}. These plots, taken from \cite{Bender:1998gh}, show the
real eigenvalues as a function of
$\varepsilon$  for the $K=2$ and $K=3$ cases of
\eq
H^{(K)}_{\varepsilon}=p^2+x^{2K}(ix)^{\varepsilon}
\qquad\quad\mbox{($K\in\NN$\,,~ $\varepsilon$ real)~,}
\label{bb1}
\en
with the implicit understanding that in each case
the quantisation contour
at $\varepsilon=0$
is the real line.
The spectrum at $K=2$, shown in figure \ref{bb12}a,
is conjectured to be real for all $\varepsilon \ge 0$, and
undergoes an extra `phase
transition' at $\varepsilon=-1$.  The real eigenvalues for
next-next-next-neighbour Stokes sectors (the $K=3$
case of (\ref{bb1})) are shown in figure \ref{bb12}b.
This time there are  phase
transitions at $\varepsilon=-2,-1$ and $0$, and the spectrum
is conjectured to be real for all $\varepsilon \ge 0$.
For {\em integer}\/ values of $\varepsilon \ge 1-K$, the reality of the
spectrum of (\ref{bb1})
was proved in \cite{Shin:2002ay}, using techniques related to those
of \cite{DDTb}. (The negative values of $\varepsilon$ covered by the
proof correspond to the extra phase transitions just mentioned.)

\begin{figure}[ht!]
\begin{center}

\includegraphics[width=0.32\linewidth]{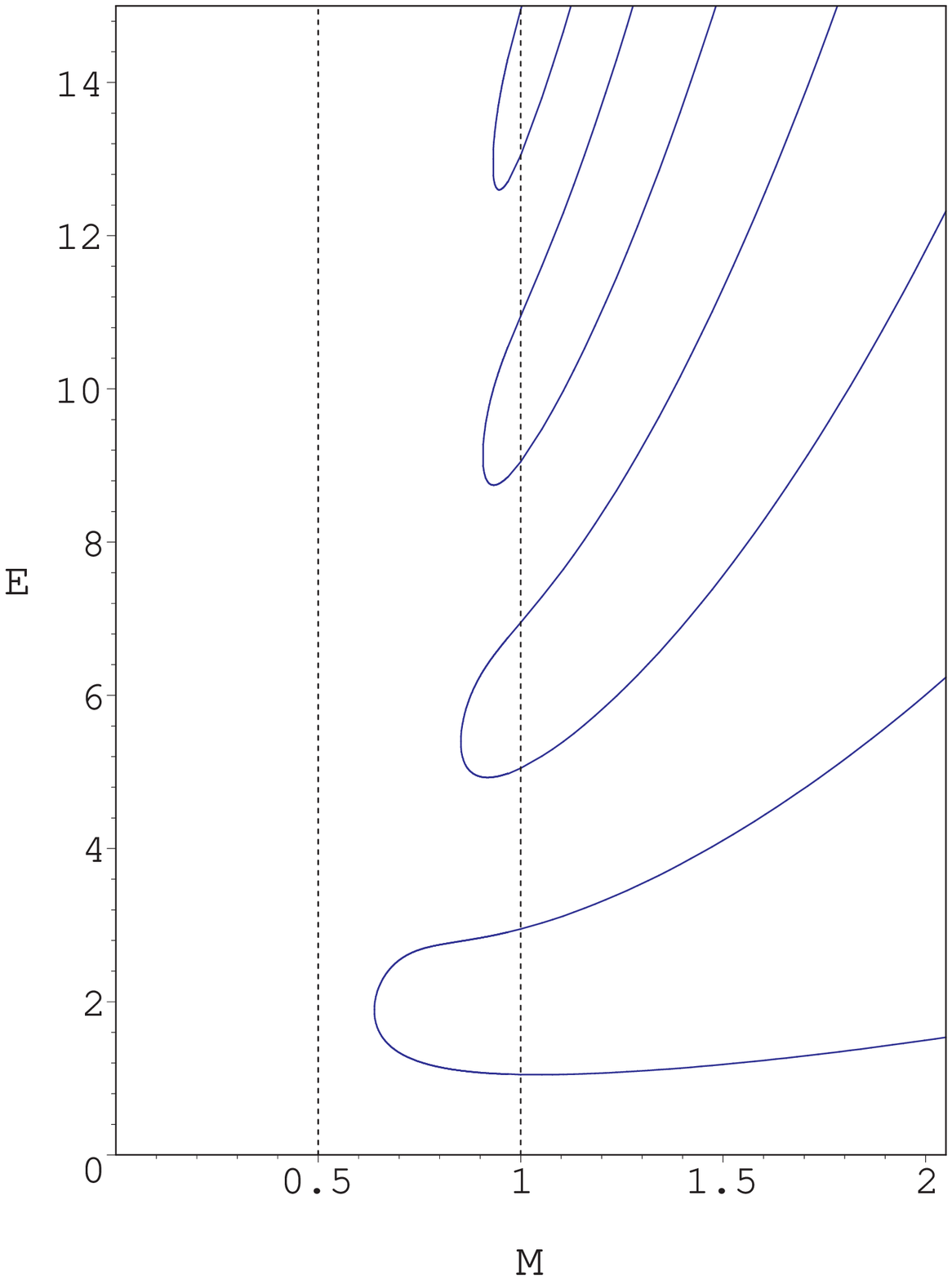}~~~~~~~~~
{}~~~~~~
\includegraphics[width=0.32\linewidth]{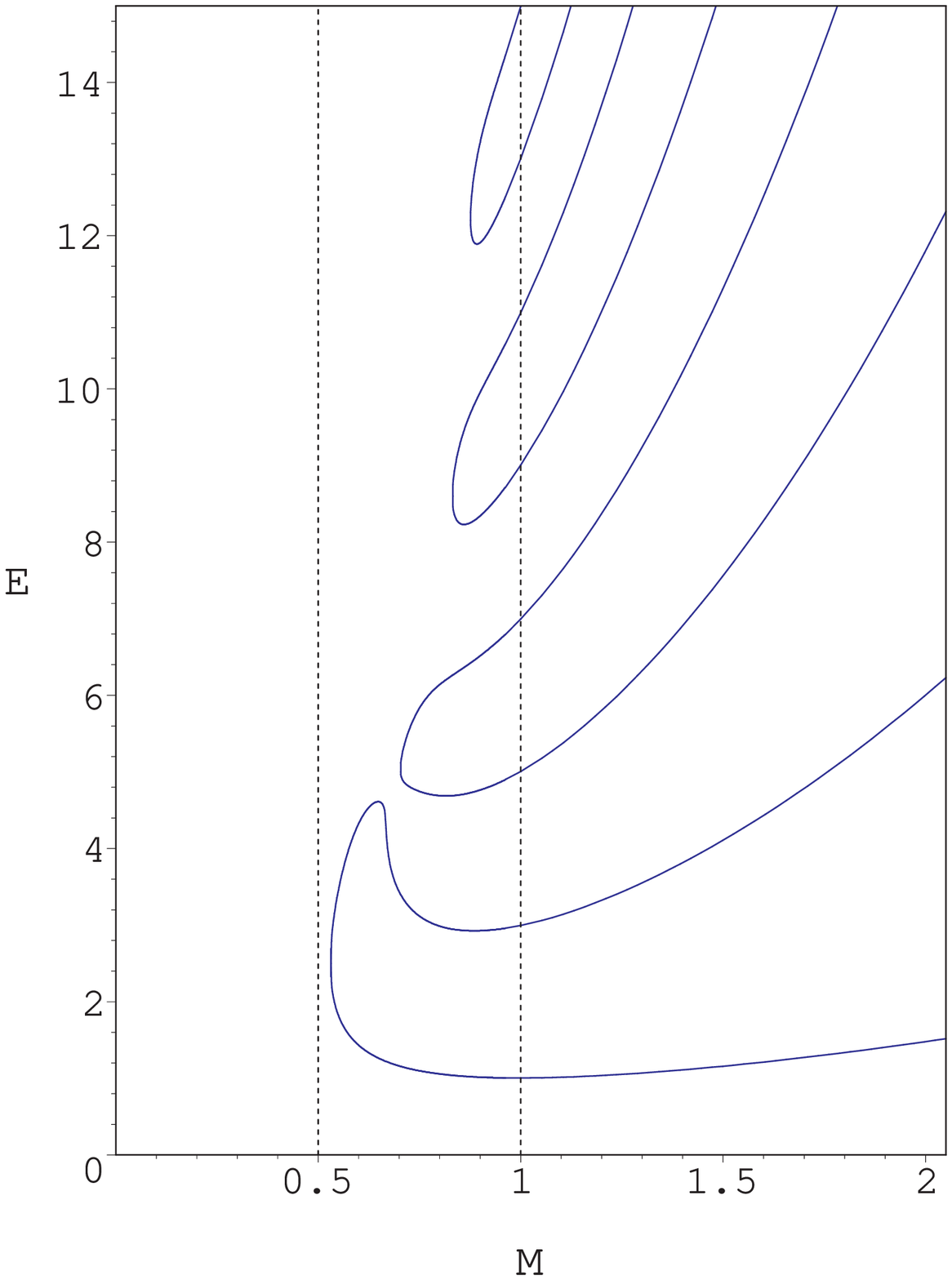}\\[1pt]
\parbox{0.33\linewidth}{~~~~~~\small\protect\ref{numerics}a: $l=-0.025$}~~~~%&
{}~~~~
\parbox{0.33\linewidth}{~~~~~~\small\protect\ref{numerics}b:
$l=-0.0025$}\\[16pt]
\includegraphics[width=0.32\linewidth]{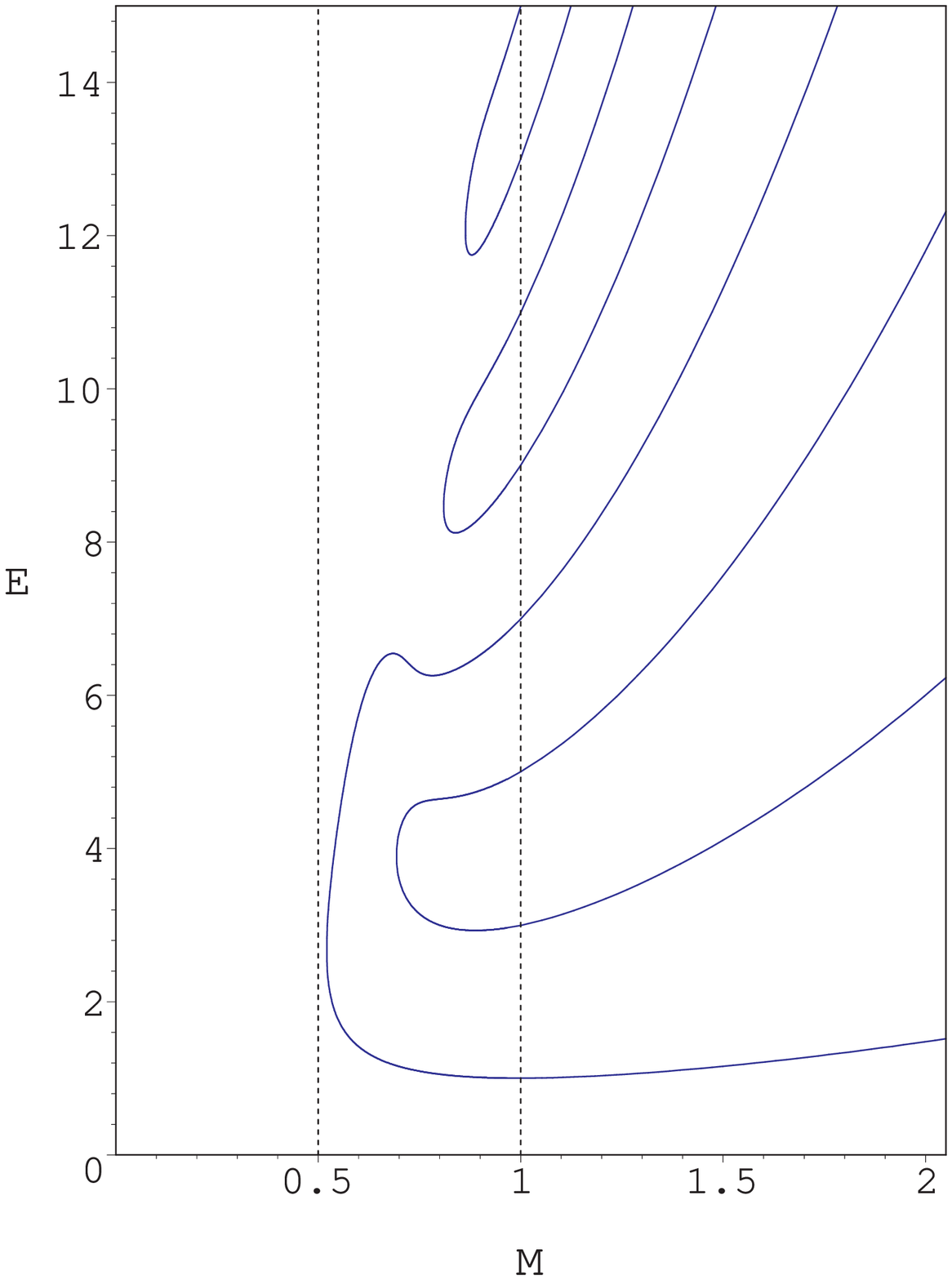}~~~~~~~~
{}~~~~~~
\includegraphics[width=0.32\linewidth]{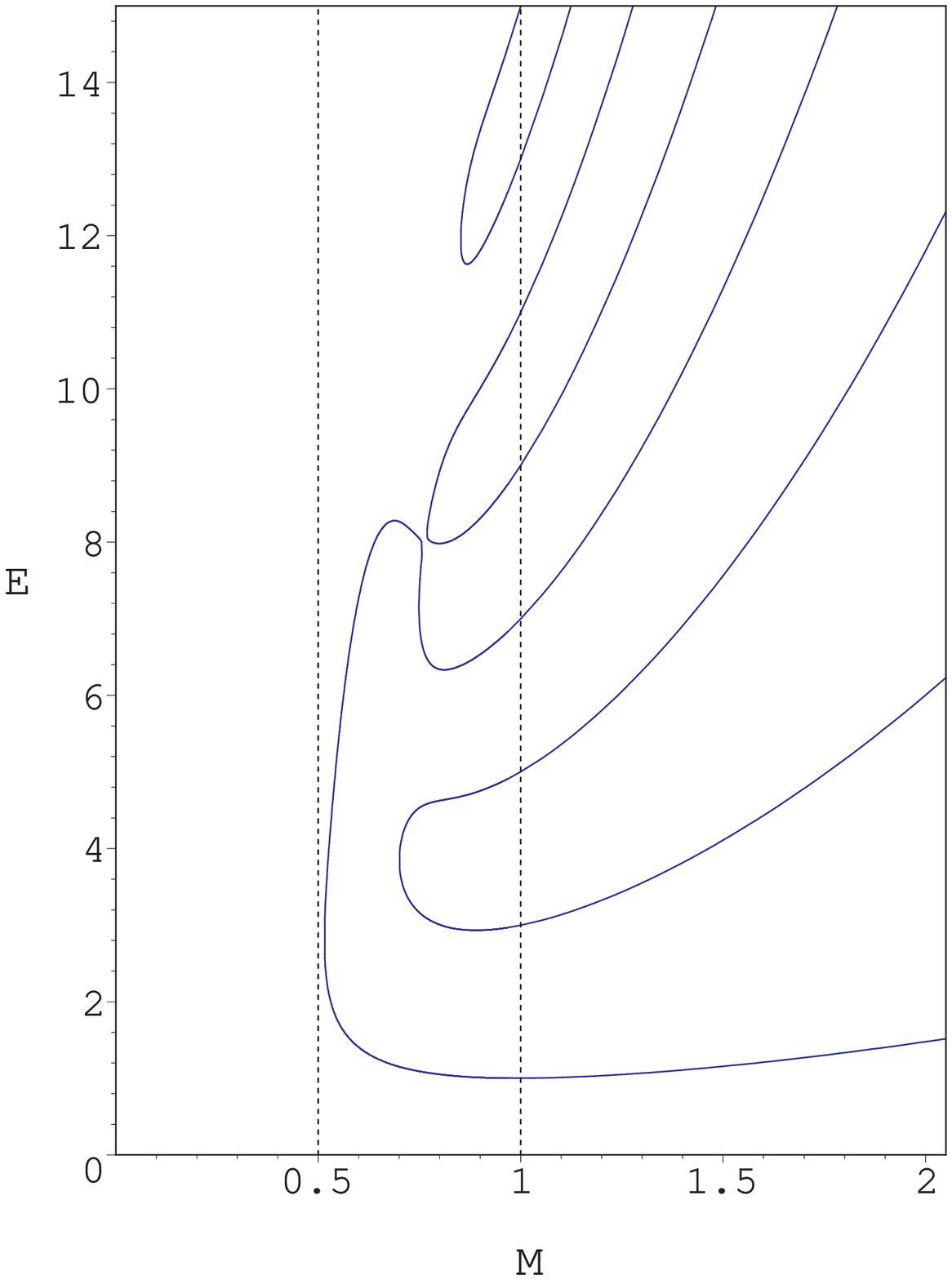}\\[1pt]
\parbox{0.33\linewidth}{~~~~~~\small\protect\ref{numerics}c:
 $l=-0.0015$}~~~~
{}~~~~
\parbox{0.33\linewidth}{~~~~~~\small\protect\ref{numerics}d:
$l=-0.001$}\\[-24pt]
\end{center}
\caption{\small  \protect{ \label{numerics}}Real eigenvalues of
$p^2-(ix)^{2M}+l(l{+}1)/x^2$ as functions of $M$.
}
\end{figure}

An alternative to looking at eigenproblems in varying pairs of
Stokes sectors is to stay with the same
Stokes sectors as for the original Bender-Boettcher problem and
explore other possibilities for the potential. In
 \cite{Dorey:1999uk} the effect of the addition of an
angular-momentum-like
term $l(l{+}1)/x^2$ was investigated, the Hamiltonian becoming
\eq
H_{M,l}= p^2-(ix)^{2M}
+ l(l+1)/x^2~.
\label{HMl}
\en
The results are shown in figure \ref{numerics}: note the completely
reversed connectivity of the real levels in figure \ref{numerics}a,
compared to that of figure \ref{f0}. The
remaining figures show how
this connectivity changes
as $l$ increases toward zero, driven by the lowest eigenvalue.
This behaviour was subsequently understood analytically, again using
tools borrowed from the world of integrable models, in~\cite{Dorey:2004fk}.

\begin{figure}[t]
\begin{center}
\includegraphics[width=0.41\linewidth]{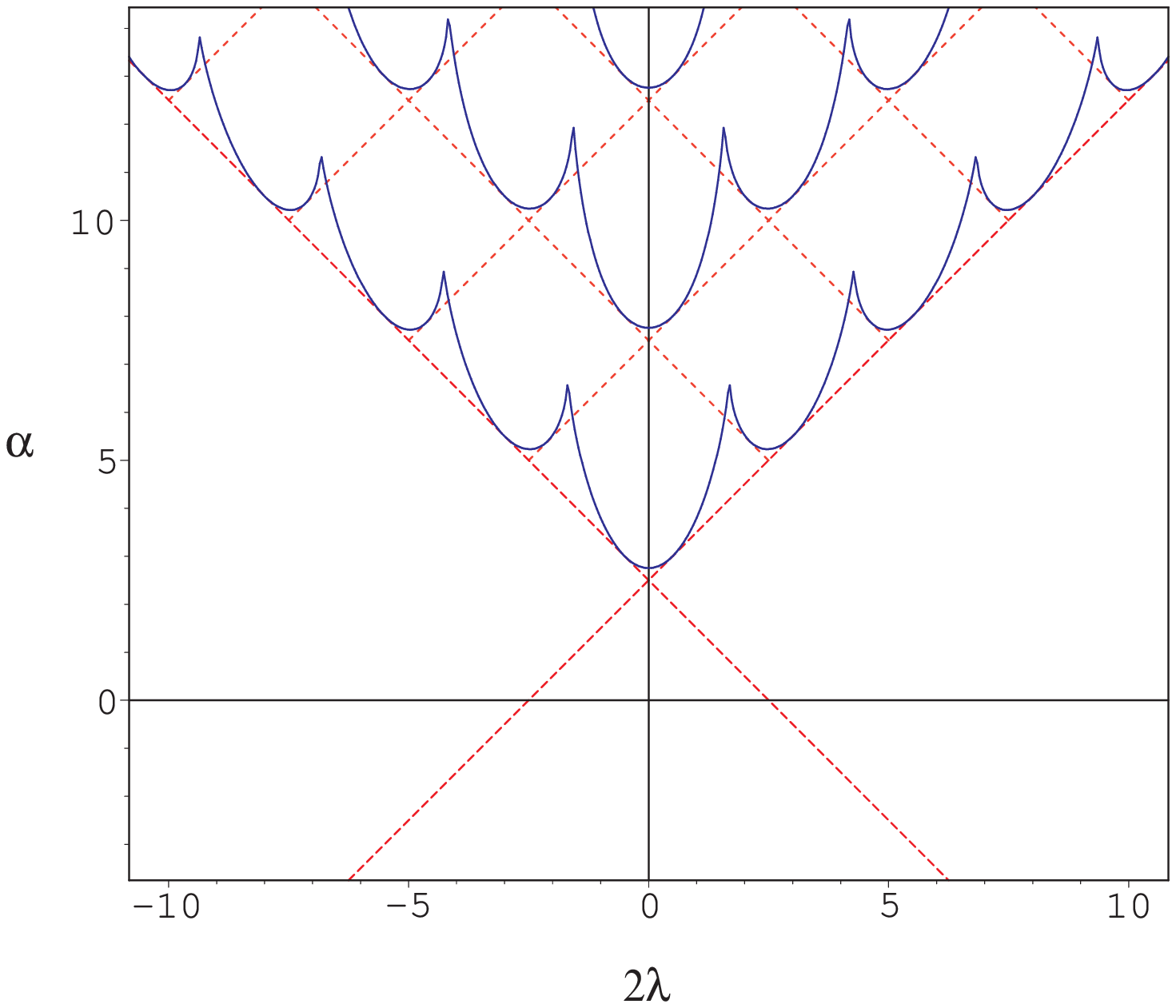}%~~~~~&~~~~~~%
{}~~~~~~~~~
\includegraphics[width=0.41\linewidth]{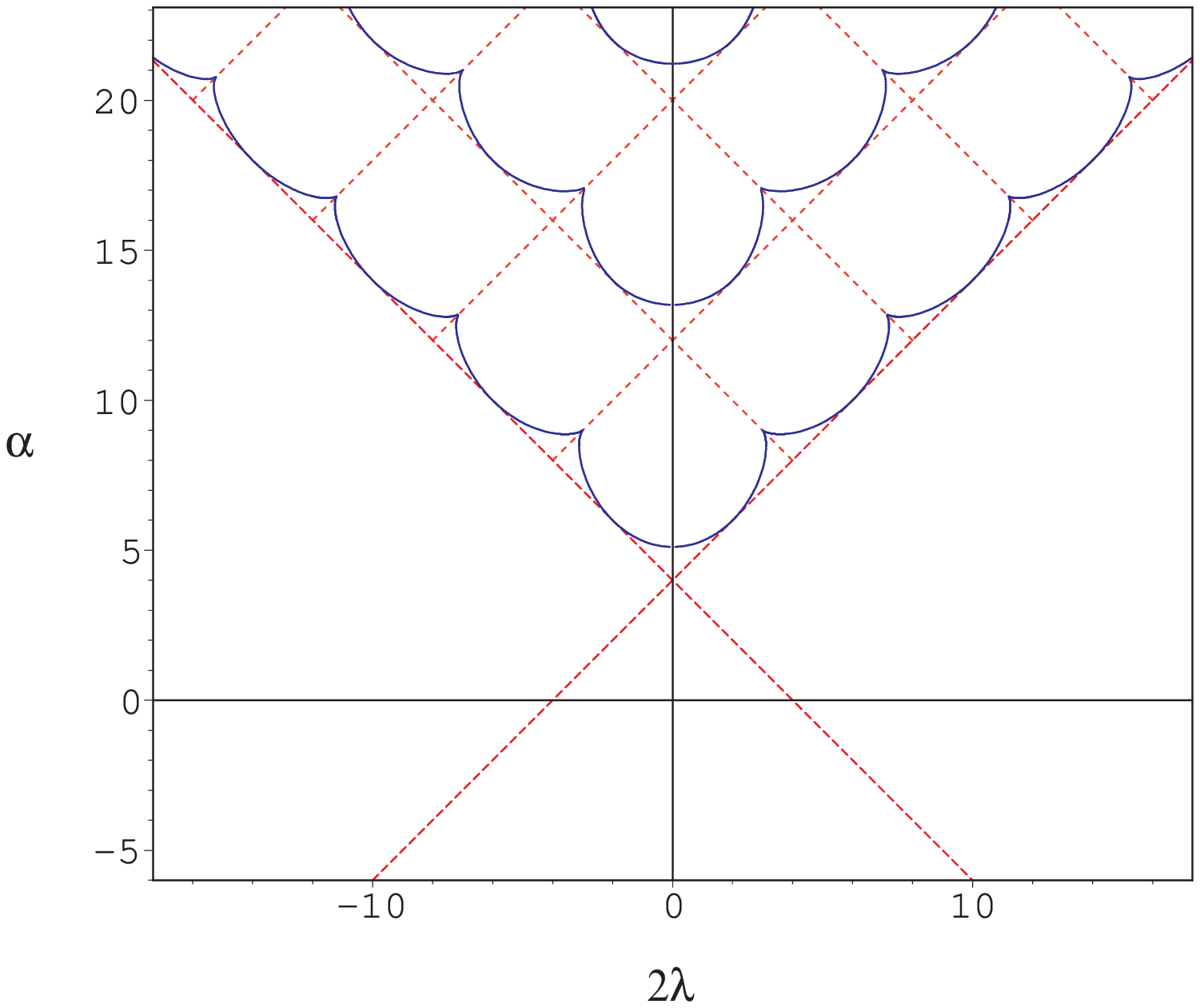}\\[1pt]
\parbox{0.33\linewidth}{~~~~~~\small\protect\ref{scans}a: $M=1.5$}~~~~%&
{}~~~~~~~~~
\parbox{0.33\linewidth}{~~~~~~\small\protect\ref{scans}b:
$M=3$}\\[22pt]
\includegraphics[width=0.41\linewidth]{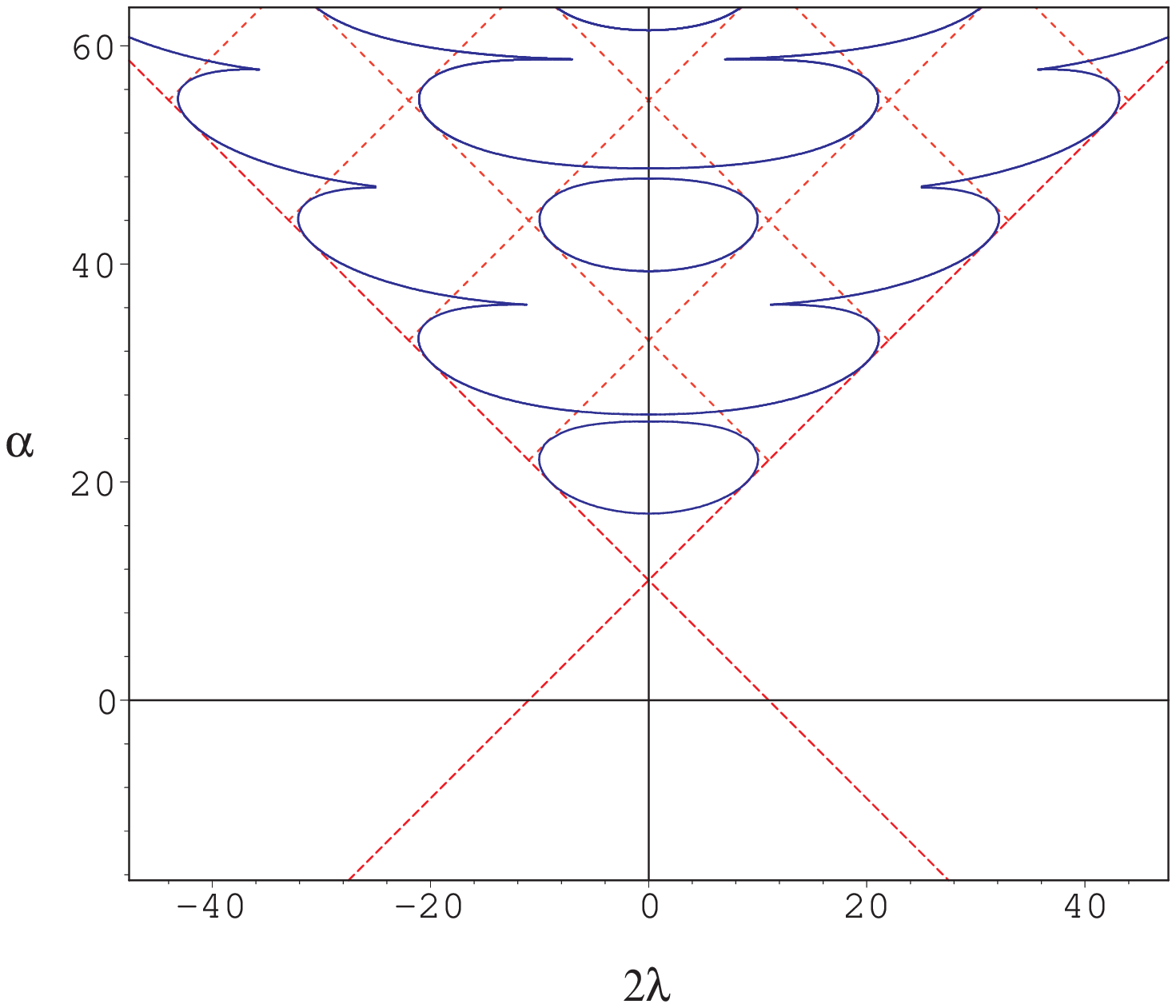}%~~~~~&~~~~~~%
{}~~~~~~~~~
\includegraphics[width=0.41\linewidth]{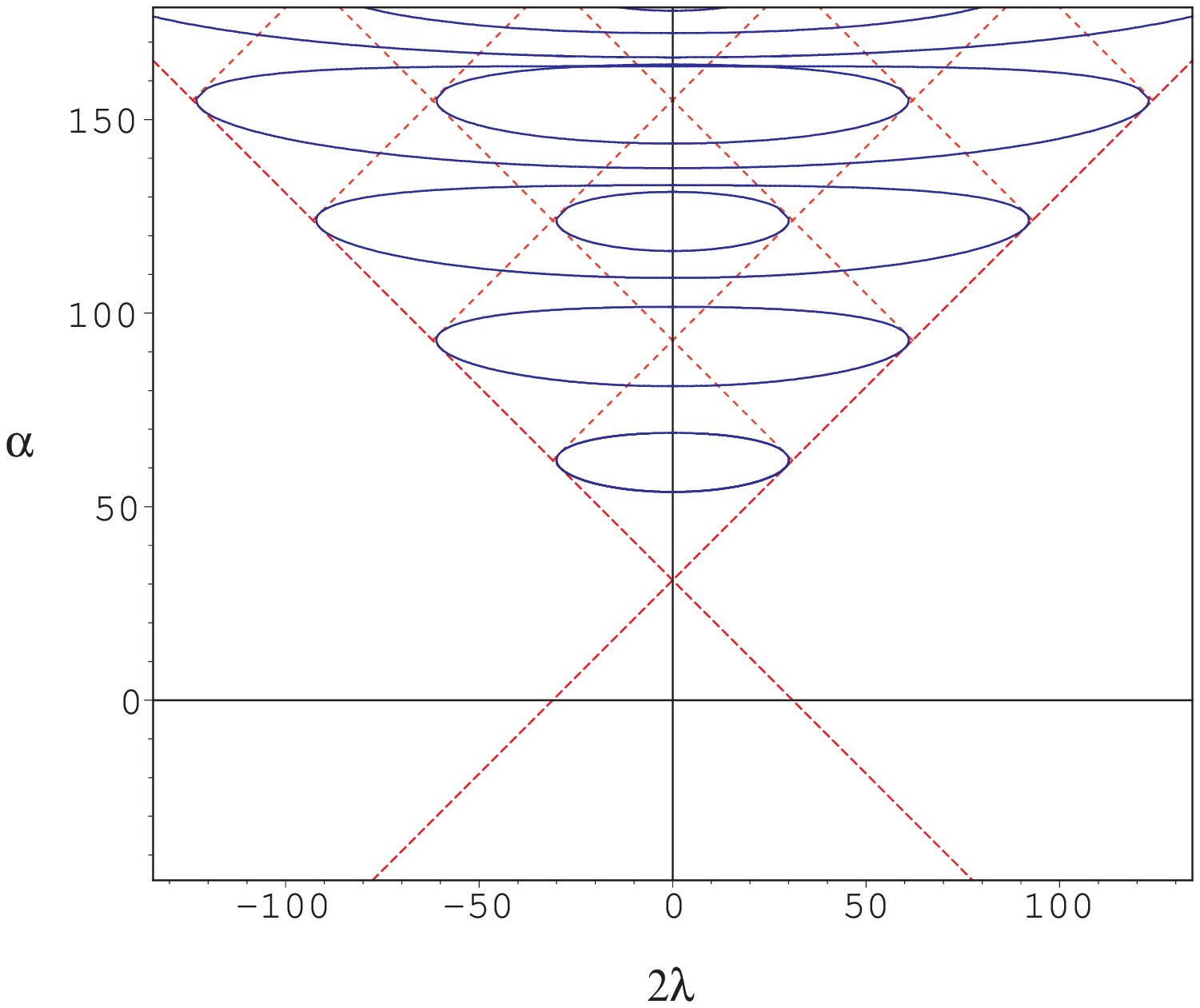}\\[1pt]
\parbox{0.33\linewidth}{~~~~~~\small\protect\ref{scans}c:
 $M=10$}~~~~
\parbox{0.33\linewidth}{~~~~~~\small\protect\ref{scans}d:
$M=30$}
\end{center}
\caption{\small \protect{ \label{scans}} Lines in the $(2\lambda,\alpha)$
plane across which complex pairs of eigenvalues appear,
for various values of $M$. In each case, the spectrum is entirely
real in the lower portion of the plot, and acquires complex
eigenvalues as the curved lines are crossed.
 }
\end{figure}

Finally, we mention one further generalisation, first studied in
detail in \cite{DDTb,Dorey:2001hi}:
 the 3-parameter family of ${\cal PT}$-symmetric Hamiltonians
  \eq
H_{M,l,\alpha}= p^2-(ix)^{2M} -\alpha(ix)^{2M-1} + l(l+1)/x^2
\en
with  boundary conditions on the Schr\"odinger-picture
wavefunctions imposed in the same
  next-nearest-neighbour pair of Stokes sectors as before.
The reality proof of \cite{DDTb} shows that
the spectrum of $\CH_{M,\alpha,l}$ is
\begin{itemize}
\item {\it real} for $M>1$ and $\alpha < M+1+|2l+1| $~;
\item {\it positive} for $M>1$ and $\alpha < M+1-|2l+1|$~.
\end{itemize}
This translates into
the spectrum being entirely real in the regions below
the dark  dashed lines (red in colour)
 $\alpha=M+1\pm2\lambda$  on the plots in
 figure \ref{scans}, where $2\lambda=2l+1$.   In fact, the full
region of spectral reality is
considerably more complicated than this initial result would suggest:
the curved cusped lines on the plots (blue in colour)
show the lines in the $(2\lambda,\alpha)$
plane across which the number of complex eigenvalues changes, for
various values of $M$. Further details will appear in \cite{soon}.

%%%%%%%%%%%%%%%%%%%%%%%%%%%%%%%%%%%%%%%%%%%%%%%%%%%%%%%%%%%
%

\subsection{Physical consistency}
Given these  surprising reality properties, a natural question
arises:  are  $\PT$-symmetric Schr\"odinger problems with entirely
real spectra such as~(\ref{cubic}) fully consistent quantum
mechanical systems?
Many recent papers~\cite{Mostafazadeh:2002hb, Bender:2002vv,
Mostafazadeh:2002id,  Bender:2004ej, Curtright:2005zk} (see also the review
articles~\cite{ Bender:2005tb,Mostafazadeh:2008pw})
have  addressed this
question.
Given a $\PT$-symmetric Hamiltonian $H$ in its unbroken $\PT$ phase,
 and denoting by  $\{ \psi_n \}$
its  set of eigenfunctions, up to a multiplicative rescaling  we have
\eq
\PT \psi_n(x)=\psi_n(x)~.
\en
We can then define  a $\PT$ inner product  as
\eq
(f,g)_{\PT}= \int_{\gamma} dx [ \PT f(x)] g(x)~,
\en
where $\gamma$ is the contour in the complex plane along which the
 wavefunction is defined.
However,  the  inner product  $(.,.)_{\PT}$ is not positive definite
 and  it is easy to check
numerically that  the orthogonality condition for the wavefunctions  is
\eq
(\psi_n, \psi_m)_{\PT} = (-1)^n  \delta_{n,m}~,~~(n,m=0,1,2,\dots)~.
\label{orto}
\en
Similarly, indirect numerical support~\cite{Bender:2000wj,
  Bender:2002eg, mez,BW} leads to
\eq
\sum_{n=0}^{\infty}   \psi_n(x) \psi_n(y)= (-1)^n \delta(x-y)~.
\label{comp}
\en
Given the  results (\ref{orto}) and (\ref{comp}) it is natural to
introduce a new  operator ${\cal C}$ such that
\eq
\psi^{\CPT}_n= \CPT \psi_n = (-1)^n \psi_n~,
\en
and define a  $\CPT$ inner product
\eq
(f,g)_{\CPT}= \int_{\gamma} dx [ \CPT f(x)] g(x)~.
\label{cpt}
\en
One can check that a $\PT$-symmetric Hamiltonian $H$ commutes with the
$\CPT$ operator.
Moreover,  $H$  is
Hermitian with respect to the $\CPT$ inner product  and it defines a unitary
time evolution. The observables correspond to operators satisfying the
following condition
\eq
O^T=
 \CPT O \, \CPT~,
\en
where $O^T$ stands for the transpose of $O$. A  major
difference between
conventional quantum mechanics and $\PT$-symmetric quantum mechanics is
that the inner product (\ref{cpt}) is generated dynamically by the
Hamiltonian:
in position space the linear operator
${\cal C}$ is given by
\eq
{\cal C}(x,y)=\sum_{n=0}^\infty \psi_n(x) \psi_n(y)~.
\en
The parity operator ${\cal P}$ in position space is
\eq
{\cal P}(x,y)=\sum_{n=0}^\infty (-1)^n \psi_n(x) \psi_n(-y)=\delta(x+y)~,
\en
and the  time reversal operator ${\cal T}$ is just complex conjugation
\eq
{\cal T}(x,y)=\sum_{n=0}^\infty (-1)^n \psi^*_n(x) \psi_n(y)~.
\en
Finally we note that further generalisations of $\PT$ symmetry have
been considered -- see for example
the $\CPT$  quantum mechanical models
discussed in~\cite{Bagchi:2004bu, Caliceti:2004xw}.

%
%%%%%%%%%%%%%%%%%%%%%%%%%%%%%%%%%%%%%%%%%%%%%%%%%%%%%%%
%
\resection{ The ODE/IM correspondence for the minimal models}
The ODE/IM correspondence~\cite{Dorey:1998pt} connects the
spectral properties of ordinary differential equations, including some
of the $\PT$-symmetric problems discussed above, with objects which
arise in the study of integrable models.
The simplest example \cite{Dorey:1998pt,Bazhanov:1998wj}
links the following $1d$  Schr\"odinger equation
\eq
\left(-\frac{d^2}{dx^2}+ (x^{2M} -E)+ \frac{l(l+1)}{x^2} \right)
\psi(x)=0~,
\label{first}
\en
where  $M>0$ and $l$ is real,
 to the conformal field theories with $c \le 1$ in the framework
proposed  by Bazhanov,
Lukyanov and Zamolodchikov~\cite{Bazhanov:1994ft, Bazhanov:1996dr}.

More precisely, equation (\ref{first}) encodes properties of the
primary operator with conformal dimension
\eq
h(M,l)=\frac{(2l
+ 1)^2 -4 M^2}{16 (M + 1)}~
\label{hhh}
\en
belonging to a conformal field theory with central charge
\eq
c(M)= 1- \frac{6 M^2}{M+1}~.
\en
Give a  pair of  coprime integers
 $\p<\q $,  the ground-state of the minimal model $\CM_{\p ,\q} $
is selected  by setting
\eq
M+1=\frac{\q}{\p}~,~~
 l+\frac{1}{2}=\frac{1}{\p}~
\en
in (\ref{first}). This corresponds to the
central charge
$
c(M)=c_{\p\q}=1-\frac{6}{\p\q}(\q{-}\p)^2~,
$
and lowest-possible conformal dimension
\eq
h(M,l)=\frac{4}{\p\q}(1-(\q{-}\p)^2)~.
\label{h}
\en
For the Ising model   $\CM_{3 ,4}$, equation (\ref{h}) gives
$h(\frac{1}{3},{-\frac{1}{6}})=0$ and
the correspondence is with the identity operator, $\One$.
For the Yang-Lee model
$\CM_{2 ,5}$, we have
$h(\frac{3}{2},{-\frac{1}{2}})=-\fract{1}{5}$ and the
correspondence  is with
the relevant operator $\phi$.
Now we observe
that the singular term   $l(l{+}1)/x^2 $ in (\ref{first})
with $l=\frac{1}{\p}-\frac{1}{2}$ can be eliminated  by a  change of
variables
\eq
x= z^{\p/2}~,~~\psi(x, E)=z^{\p/4-1/2} y(z,E)~.
\label{changev}
\en
After rescaling $z\to (2/\p)^{2/\q}z $ equation (\ref{first})  finally becomes
\eq
\left(-\frac{d^2}{dz^2}+
z^{\p-2} (z^{\q-\p} -\tilde E) \right) y(z,\tilde E)=0~,
\label{trf}
\en
where
\eq
\tilde E = \left(\frac{\p}{2}\right)^{2-2\p/\q}E~.
\en
The change of variable (\ref{changev})  has replaced a singular
potential defined on a
multi-sheeted Riemann surface by a simple polynomial. Therefore
any solution
to the transformed equation~(\ref{trf}) is automatically
single-valued  around $z=0 $.

Zero monodromy conditions have already  played a central r\^ole in the
ODE/IM correspondence~\cite{Bazhanov:2003ni,Feigin:2007mr}, and
therefore
it is natural to explore the constraints that this
property imposes on the constant $l$ in (\ref{first}) and (\ref{hhh})
\cite{Dorey:2007ti}.

%
%
%%%%%%%%%%%%%%%%%%%%%%%%%%%%%%%%%%%%%%%%%%%%%%%%%%%%%%%%%
%
\subsection{ Monodromy properties  and the Kac table}
\begin{figure}[ht]
\begin{center}
\epsfxsize=.5\linewidth\epsfbox{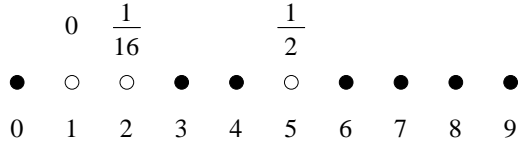}
\end{center}
\caption{\small The holes  in the infinite sequence of
integers  for
the critical Ising model $\CM_{3,4}$. The holes are
at $1$, $2$ and $5$. } \label{fig13}
\end{figure}

\begin{figure}[ht]
\begin{center}
\epsfxsize=.5\linewidth\epsfbox{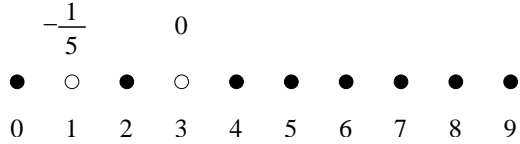}
\end{center}
 \caption{\small   Holes for the Yang-Lee model
$\CM_{2 ,5}$\,, at $1$ and $3$. }\label{fig14}
\end{figure}

To see which other primary states  have
similarly-trivial monodromy,  start from (\ref{first}) with generic   $l>-1/2 $ and perform
the  transformation (\ref{changev}). The result is
\eq
\left(-\frac{d^2}{dz^2}+  \frac{\tilde l  (\tilde l +1)}{z^2}
+ z^{\p-2} (z^{\q-\p} -\tilde E)
 \right) y(z,\tilde E,\tilde l)=0~,
\label{singe}
\en
where $
2(\tilde{l}+\fract{1}{2})=\p (l+ \fract{1}{2})$.
The  presence of the Fuchsian singularity at $z=0 $, compared to
(\ref{trf}),   means that the zero monodromy condition of a generic
solution is no longer guaranteed.
In general, equation (\ref{singe})  admits a pair of solutions
\eq
\chi_1(z)=z^{\lambda_1}\sum_{n=0}^\infty c_n z^n~;~~~~
\chi_2(z)=z^{\lambda_2}\sum_{n=0}^\infty d_n z^n~~,
\en
where $\lambda_1=\tilde{l} +1 >  \lambda_2=-\tilde{l} $
are the two roots of the indicial equation
$ \lambda(\lambda-1)-\tilde l(\tilde l+1)=0~$
and
\eq
\chi_j(e^{2 \pi i}z) =e^{2 \pi i \lambda_j} \chi_j(z)~,~~j=1,2~.
\en
%
%
%%%%%%%%%%%%%%%%%%%%%%%%%%%%%%%%%%%%%%%%%%%%%%%%%%%%55
%
%
Therefore, the  general solution is a linear combination
\eq
y(z,\tilde E,\tilde l)=\sigma\chi_1(z)+\tau \chi_2(z)
\en
and we  shall demand that the monodromy
of $y(z) $  around $z=0$
is projectively trivial: $y(e^{2 \pi i}z) \propto y(z)$.
We shall see that this  condition imposes
\begin{itemize}
\item[i)] $2\tilde l+1 $ is a positive integer;
\item[ii)] The allowed values of  $2\tilde l+1 $  form  the
set of holes of the infinite sequence
\eq
\p\,r+\q\,s~~,~~r\,,s=0,1,2,3\dots.
\label{rep}
\en
We call the set of integers (\ref{rep})
`representable', and denote them  by $\Rth_{\p\q} $.
\end{itemize}
As a consequence of i) and ii), it turns out that
the allowed conformal dimensions
\eq
h(M,l)=\frac{(2\tilde l+1)^2-(\p-\q)^2}{4\p\q}
\en
exactly reproduce the set of conformal weights of the primary
operators  in the Kac table of  $\CM_{\p ,\q} $. Figures~\ref{fig13}
and \ref{fig14} illustrate the situation for the Ising and the Yang-Lee
models.
%
%
%%%%%%%%%%%%%%%%%%%%%%%%%%%%%%%%%%%%%%%%%%%%%%%%%%%%%%%%%%%%%%%%%%%
%
%
%
%
%
%
%%%%%%%%%%%%%%%%%%%%%%%%%%%%%%%%%%5
%

To establish these claims, we first notice that the requirement that
the general solution $y(z)$ be projectively trivial
means that $\chi_1(z) $ and $\chi_2(z) $ must have the same monodromy.
This  implies
\eq
\lambda_1-\lambda_2=2\tilde l+1\in\NN\,,
\en
and therefore restricts $\tilde l $ to be an integer or half integer.
However, in such a circumstance, while $\chi_1(z) $
keeps its power series expansion, $\chi_2(z) $
generally acquires a logarithmic contribution
\eq
\chi_2(z)= D\chi_1(z)\log(z)+\frac1{z^{\tilde l}}
\sum_{n=0}^\infty d_nz^n~,
\label{chilog}
\en
and unless $D=0 $, this will spoil the projectively trivial monodromy of
$y(z)$.
%
%
%%%%%%%%%%%%%%%%%%%%%%%%%%%%%%%%%%%%%%%%%%55
%
The log term is
absent if and only if the  recursion relation for the $d_n $'s with $D=0 $
\eq
n\,(n-2\tilde l-1)\,d_n=d_{n-\q}-\tilde E\,d_{n-\p}
\en
with the initial conditions $d_0=1 $, $d_{m<0}=0 $
admits a solution.

Consider first the case  $2  \tilde l+1 \notin \Rth_{\p\q} $.
Starting from the given initial conditions, the recursion
relation  generates a solution of the form
\eq
\chi_2(z)= \frac{1}{z^{\tilde l}} \sum_{n=0}^{\infty} d_{n}  z^{n}
\en
where the only nonzero $d_n$'s are those
for which the label $n$ lies in the set $\Rth_{\p\q} $.
Given that $2\tilde l+1\notin \Rth_{\p\q} $, for these values of $n$
the factor $n(n-2\tilde l-1) $
on the LHS of the recursion relation is never
zero, and the procedure is well-defined.
%
%
%%%%%%%%%%%%%%%%%%%%%%%%%%%%%%%%%%%%%%%%%%%%%%%%%%%%%%%%%%%%%%%%%%%%%%%
%
If instead $2\tilde l+1\in\Rth_{\p\q} $, then the recursion equation
taken at $n=2\tilde l+1 $
yields the additional condition
\eq
\tilde E\,d_{2 \tilde l+1 -\p}-d_{2 \tilde l +1 -\q}=0~,
\en
which is inconsistent for generic $\tilde E $, and so the
log term is required.
If the coprime integers $\p $ and $\q $ are larger than 1 then
the number of   `nonrepresentable' integers $\NN_{\p\q}$  with
$\ZZ^+=\Rth_{\p\q}\cup\NN_{\p\q}$ is in fact
$|\NN_{\p\q}|=\frac{1}{2}(\p{-}1)(\q{-}1) $, a result which goes back
to Sylvester~\cite{sylv} and matches the number of primary fields in the Kac table.
\resection{ Applications to condensed-matter physics}

There are many  interesting potential physical applications of
$\PT$-symmetry,  ranging from  optics
to the more  speculative proposals
related to conformal gravity~\cite{Bender:2008vh},
and Higgs phenomena in the Standard Model of particle
physics~\cite{Bender:1999ek}.
A partial list of examples can be found in the
review~\cite{Mostafazadeh:2008pw} on Pseudo-Hermitian quantum
mechanics.  An `experimental' means of distinguishing between a
(Dirac) Hermitian and a $\PT$-symmetric Hamiltonian  through the
quantum brachistochrone is an interesting recent development
\cite{Bender:2006fe, Assis:2007dh}. In section~\ref{brpoly} a general
relationship between
the Yang-Lee model and non-intersecting branched-polymers was
discussed.  Here we would like to report   on a  simple
application
of  $\PT$-symmetric quantum mechanics to the ferromagnetic Ising system
and briefly mention
a surprising correspondence between the equation~(\ref{first}) and
experiments in cold atom physics.

Recently  Lamacraft and  Fendley~\cite{Lamacraft:2008:fd} were able to
connect the study of universal amplitude ratios (see also
~\cite{Salas:1999qh})
in the quantum spin-chain version of the  Ising  model to a simple
$\PT$-symmetric Schr\"odinger problem. The
Hamiltonian of the quantum spin-chain   Ising model is
\eq
\hat{{\cal H}}[{\bf \sigma},\text{h}]= - \sum_{i=1}^{L} [ \text{h} \sigma_i^x+ \sigma_i^z \sigma_{i+1}^z]~,~~~~~(\sigma^z_{L+1}=\sigma_1^z)~,
\label{isingchain}
\en
where $\sigma^x$ and $\sigma^z$ are  Pauli matrices.
The model~(\ref{isingchain}) is critical at  $\text{h}=\text{h}_c=1$ and
in its ordered and disordered phases for   $\text{h} <1$
and $\text{h}>1$  respectively.
The magnetization operator  $\hat{\text{M}}=\sum_i \sigma_i^z/2$
does not commute with $\hat{{\cal H}}$. Setting $\text{h}=\text{h}_c$ and considering the
continuum limit versions of the operators  $\hat{{\cal H}}$ and
$\hat{\text{M}}=\int_0^{L} dx \, \sigma(x)$, we can
define the  magnetisation distribution in the ground-state $|0 \rangle$ as
\eq
P(m)= \langle 0| \delta(m-\hat{\text{M}})|0 \rangle = \int_{-\infty}^{\infty} \frac{d\lambda}{2\pi} e^{-i \lambda m}  \chi(\lambda)~,
\en
where
\eq
\chi(\lambda) = \langle 0|e^{i \lambda \hat{\text{M}}}|0 \rangle~
\en
is the generating function of the moments of the distribution $P(m)$
\eq
 \langle 0| \hat{\text{M}}^{2n} |0 \rangle = \langle m^{2n} \rangle= \int_{-\infty}^{\infty} m^{2n} P(m) \, dm = (-1)^n \left [\frac{d}{d \lambda^{2n}} \chi(\lambda)\right]_{\lambda=0}~.
\en
It is a direct consequence of the Yang and Lee results~\cite{Yang:1952be, Lee:1952ig} that the function  $\chi(\lambda)$  is an entire function of $\lambda$, and a consequence of the Hadamard factorisation theorem that
\eq
\chi(\lambda) = \prod_{n=0}^{\infty} \left( 1-  \frac{ \lambda^2}{ E_n} \right)~.
\en
Finally    $\chi$ can be  related to the
spectral determinant $T$~\cite{Dorey:1999uk} of   the   following  $\PT$-symmetric Hamiltonian (cf. (\ref{HMl}))
\eq
H_{7,-\fract{1}{2}}= p^2-(ix)^{14} -  \frac{1}{4x^2}~.
\en
The relation is simply
\eq
\chi(\lambda)=\frac{1}{2} T(E)~,~~~~~ \lambda^2 \propto E~.
\en

The second application,
proposed by Gritsev et al.\ in~\cite{GADP},
is  related to  interference experiments in
cold atom physics.
In a typical interference experiment \cite{GADP,IGD,GMMPS}, a pair of
parallel $1d$ condensates of length $L$ is created using standard
cooling
techniques and an highly anisotropic double well radio-frequency-induced
 micro-trap.
 After the condensates are released from the trap, interference
 fringes are observed using a laser beam. If the
 condensate has the form of two coaxial rings~\cite{GMMPS},  the
 associated `full distribution function' in the zero temperature limit can be mapped,
 through the ODE/IM correspondence, to the spectral
 determinant of the  Schr\"odinger equation~(\ref{first}).

\resection{Conclusions}
In this short review we have discussed non-Hermitian quantum mechanics,
integrable quantum  field theory, links between them via an ODE/IM
correspondence and some  applications
to condensed-matter physics. The material is by no means exhaustive
but rather reflects the authors' experience and interests.
Apart from a small number of exceptions,
most of the
papers on $\PT$-symmetry concern one-dimensional Schr\"odinger
problems  in the
complex domain;  many fundamental issues related to their full
consistency as quantum mechanical models
have been successfully addressed.
Since the early 1990's two-dimensional  relativistic non-unitary quantum field theory,
particularly the
$2d$ scaling Yang-Lee model, has been extensively studied by
the integrable model community.
Considering
the relatively-restricted physical relevance usually  given to  these
models, a large quantity of exact results have been
produced over the years.  However, none of the
fundamental interpretation issues, of great concern in the
$\PT$-symmetry community, have been completely clarified. We think
this in an interesting open problem which deserves attention.

\medskip

\noindent{\bf Acknowledgements --}
RT thanks the organizers of the conference
``Non-Hermitian Hamiltonians in Quantum Physics"  for the  Homi Bhabha
Centenary celebrations in  Mumbai for  the invitation to speak.
We are grateful to  Ferdinando Gliozzi, Anna Lishman,  Adam Millican-Slater,   Junij Suzuki and G\'erard Watts for collaboration
over the years on some of the material reported in this review.
We also
thank G\'erard Watts for the TCSA data points used in
figures \ref{fig4} and \ref{fig5}, Carl Bender and Stefan Boettcher
for permission to use the plots shown in figure \ref{bb12},
and Ingo Runkel, G\'erard Watts and Robert Weston for discussions.
This project was  partially supported by  grants  from the
 Leverhulme Trust and from the  INFN grant PI11.

%
%%%%%%%%%%%%%%%%%%%%%%%%%%%%%%%%%%%%%%%%%%%%%
%
%

%
\end{document}